\documentclass[]{jfm}

\usepackage{graphicx}
\usepackage{newtxtext}
\usepackage{newtxmath}
\usepackage{natbib}
\usepackage{hyperref}
\usepackage{mathtools}
\usepackage{amsfonts}
\usepackage{siunitx}
\usepackage{booktabs}

\hypersetup{
    colorlinks = true,
    urlcolor   = blue,
    citecolor  = black,
}

\newcommand{\RomanNumeralCaps}[1]

\DeclareMathOperator{\tr}{tr}

\newcommand{\D}{\mathrm{D}}
\newcommand{\bT}{\boldsymbol{\mathrm{T}}}
\newcommand{\bD}{\boldsymbol{\mathrm{D}}}
\newcommand{\bI}{\boldsymbol{\mathrm{I}}}
\newcommand{\bS}{\boldsymbol{\mathrm{S}}}
\newcommand{\bsigma}{\boldsymbol{\sigma}}
\newcommand{\btau}{\boldsymbol{\tau}}
\newcommand{\bt}{\boldsymbol{t}}
\newcommand{\dd}{\mathrm{d}}
\newcommand{\bu}{\boldsymbol{u}}
\newcommand{\RR}{\mathbb{R}}

\newcommand{\bc}{\boldsymbol{c}}
\newcommand{\bff}{\boldsymbol{f}}
\newcommand{\br}{\boldsymbol{r}}
\newcommand{\bn}{\boldsymbol{n}}

\newcommand{\Ac}{\mathcal{A}}
\newcommand{\Cc}{\mathcal{C}}
\newcommand{\Fc}{\mathcal{F}}
\newcommand{\Jc}{\mathcal{J}}

\numberwithin{equation}{section}

\title{Modeling sea ice in the marginal ice zone as a dense granular flow with rheology inferred from discrete element model data}

\author{Gonzalo G. de Diego \aff{1}
  \corresp{\email{gg2924@nyu.edu}},
  Mukund Gupta \aff{2},
  Skylar A. Gering \aff{3},
  Rohaiz Harris \aff{3},
 \and Georg Stadler \aff{1}}

\affiliation{\aff{1} Courant Institute, New York University, New York, USA,
\aff{2} Department of Geoscience and Remote Sensing, Delft Institute of Technology, The Netherlands,
\aff{3} Environmental Science and Engineering, California Institute of Technology, California, USA
}

\begin{document}
\maketitle

\begin{abstract}
	The marginal ice zone (MIZ) represents the periphery of the sea ice cover. In this region, the macroscale behavior of the sea ice results from collisions and enduring contact between ice floes. This configuration closely resembles that of dense granular flows, which have been modeled successfully with the $\mu(I)$ rheology. Here, we present a continuum model based on the $\mu(I)$ rheology which treats sea ice as a compressible fluid, with the local sea ice concentration given by a dilatancy function $\Phi(I)$. We infer expressions for $\mu(I)$ and $\Phi(I)$ by nonlinear regression using data produced with a discrete element method (DEM) which considers polygonal-shaped ice floes. We do this by driving the sea ice with a one-dimensional shearing ocean current. The resulting continuum model is a nonlinear system of equations with the sea ice velocity, local concentration, and pressure as unknowns. The rheology is given by the sum of a plastic and a viscous term. In the context of a periodic patch of ocean, which is effectively a one dimensional problem, and under steady conditions, we prove this system to be well-posed, present a numerical algorithm for solving it, and compare its solutions to those of the DEM. These comparisons demonstrate the continuum model's ability to capture most of the DEM's results accurately. The continuum model is particularly accurate for ocean currents faster than 0.25 m/s; however, for low concentrations and slow ocean currents, the continuum model is less effective in capturing the DEM results. In the latter case, the lack of accuracy of the continuum model is found to be accompanied by the breakdown of a balance between the average shear stress and the integrated ocean drag extracted from the DEM. Since this balance is expected to hold independently of our choice of rheology, this finding indicates that continuum models might not be able to describe sea ice dynamics for low concentrations and slow ocean currents.
\end{abstract}


\section{Introduction}\label{sec:intro}

The periphery of the ice cover is known as the marginal ice zone (MIZ) and consists of relatively small, typically polygon-shaped ice floes. It is often defined as the region where ocean waves play an important role in shaping the morphological properties of the ice \citep{dumont2022}. On large scales, sea ice dynamics are typically described with Hibler's model \citep{hibler1979}, which treats ice as a viscoplastic fluid whose yield strength depends on the sea ice concentration and thickness. This model was developed for the central ice pack, where ice floes are closely interlocked and deformation is mostly due to the opening of leads or the formation of ridges. Recently, elasto-brittle rheologies have also been used to model the evolution of the central ice pack. This class of rheological models, which were first proposed by \citet{girard2011}, appear to be superior to Hibler's model in capturing the ice deformation field \citep{rampal2019}.

In the MIZ, however, it is the collisions and enduring contact between ice floes that give rise to the macroscale dynamical properties of the ice cover \citep{feltham2005, herman2011, herman2022}. This configuration closely resembles that of dense granular flows, albeit at different spatial scales, since practically all studies for granular materials consider e.g.~polystyrene beads, glass beads, and sand, whose particles’ diameters are in the order of 0.1 and 1 mm \citep{gdr2004}. Dense granular flows have been successfully modeled with the so-called $\mu(I)$ rheology \citep{dacruz2005}. The dense granular flow regime is understood as a transition between the quasi-static and dilute flow regimes. Whenever grain inertia is negligible, a quasi-static regime emerges which is often modeled as an elastoplastic solid \citep{nedderman1992}. The critical state at which plastic deformation occurs is characterized with a Coulomb-like criterion dependent on a so-called internal angle of friction \citep{wood1990}. Conversely, under great agitation and/or dilute concentrations of grains, particles interact only through binary, instantaneous, uncorrelated collisions. As a result, ideas from kinetic theory become applicable in this dilute regime \citep{jenkins1983}. However, in dense granular flows, grains interact via collisions and enduring contacts, such that inertial effects are important yet the collisions may no longer be assumed to be binary, instantaneous, or uncorrelated in general. This transitional regime is characterized in terms of the inertial number $I$ and an effective friction coefficient $\mu$ dependent on $I$ \citep{dacruz2005}.

Existing models for the MIZ recognize the importance of both collisions and plastic deformation, and derive rheological models based on first principles \citep{shen1987, gutfraind1997,   feltham2005}. Recently, \cite{herman2022} suggests the use of the $\mu(I)$ rheology for modeling sea ice in the MIZ and derives a $\mu(I)$ function from computations performed with a discrete element method (DEM). In these computations, disk-shaped ice floes are sheared by a moving wall in the classical manner of rheological studies. Unlike the previous models for the MIZ, \cite{herman2022} infers the rheological properties from data generated by a DEM. In particular, \cite{herman2022} fits a $\mu(I)$ function to the DEM’s data, although the resulting continuum model and its accuracy in replicating the DEM’s results is not examined.

This work represents an advance in the development of a continuum model for the MIZ that could improve the accuracy of Hibler's model, which is currently used in large-scale climate models over the MIZ, see for example \citep{cesm2}. A comparison of Hibler's model with the DEM data is presented in section \ref{subsec:hibler}, where it is demonstrated that it cannot capture the DEM results accurately. We extend the investigation initiated in \cite{herman2022} and explore the $\mu(I)$ rheology's accuracy in modeling sea ice dynamics in the MIZ. We infer a $\mu(I)$ function from data produced with the DEM implemented in SubZero \citep{manucharyan2022}. This DEM considers polygon-shaped ice floes that are driven by oceanic currents in an open patch of ocean, a setup which we believe to be more natural for studying sea ice than the classical shearing test with a moving wall and disk-shaped ice floes. This inference results in a continuum viscous fluid model whose rheology is given by the sum of a viscous and a plastic term. Moreover, for this system to be well-posed, the emerging model problem requires the continuum to be compressible and complemented with a constraint on the global sea ice concentration. Assuming the continuum to be compressible requires the inference of a dilatancy function $\Phi(I)$ from the DEM computations which establishes a relationship between local sea ice concentration and the inertial number $I$. 

The contributions of this paper can be summarized as follows: (1) Inference of the $\mu(I)$ and $\Phi(I)$ constitutive functions for sea ice in the MIZ from data produced with the DEM. These computations are performed in an open ocean configuration where the sea ice is sheared by ocean currents. (2) Analysis of the resulting continuum model, establishing the existence and uniqueness of solutions. (3) Determination of the continuum model's range of validity by comparing its numerical solutions to those of the DEM.

We remark that the analysis of the continuum model and its comparisons with the DEM are restricted to a steady one dimensional setup. The model can be extended to unsteady two dimensional problems as explained in section \ref{subsec:2D}, although we expect new complications will arise with these extensions. For example, \cite{barker2015} demonstrate the emergence of time-dependent instabilities in $\mu(I)$ models, which \citet{schaeffer2019} remedy with further modifications of the model. These potential complications should be studied carefully in future investigations.

This paper is structured as follows. In section \ref{sec:formulation}, we formulate the continuum model, first in a general two-dimensional unsteady setting, then in the one dimensional steady configuration considered in this paper. In this formulation, two functions, $\mu(I)$ and $\Phi(I)$, are to be inferred from DEM data. This inference is presented in section \ref{sec:inferring}. Section \ref{sec:analysis} contains a detailed analysis of the continuum model resulting from this inference. This analysis examines several properties of the momentum equation, the numerical solution of the continuum model, and its well-posedness. Then, in section \ref{sec:comparison_DEM}, we compare the continuum model and the DEM. This comparison allows us to establish the range of validity of the continuum model and its limitations. In section \ref{sec:similarities_model}, we discuss the similarities and differences between our continuum model and other sea ice models, such as Hibler's model. We then end this paper with section \ref{sec:conclusions}, where we recommend potential extensions of this work to be explored in the future.

\section{Mathematical formulation of the continuum problem}\label{sec:formulation}

The dense flow regime represents a transition between the quasi-static and dilute flow regimes \citep{dacruz2005}. This transitional regime is characterized in terms of the inertial number $I$, an effective friction coefficient $\mu(I)$, and, whenever the continuum is assumed to be compressible, a dilatancy function $\Phi(I)$. Below, we define these three terms and present a general formulation of the $\mu(I)$ rheology in two dimensions and in the one dimensional steady configuration considered in the subsequent sections of this paper.

\subsection{The two-dimensional setting}\label{subsec:2D}

Although the problems presented in this paper are effectively one dimensional, we first present the general form of the flow model in two dimensions for completeness. We denote the ice velocity, concentration, and Cauchy stress tensor by $\bu$, $A$, and $\bsigma$, respectively. We write the components of the Cauchy stress tensor and the velocity vector field as 
\begin{align}
	\bsigma = \left[ \begin{array}{cc}
		\sigma_{xx} & \sigma_{xy} \\
		\sigma_{xy} & \sigma_{yy}
	\end{array} \right] \quad \text{and} \quad \bu = (u,v),
\end{align}
respectively. We assume that the morphology of the ice floes remains invariant by neglecting all thermomechanical effects, such as fracturing, melting, or ridging, that can change the shape of a floe. For simplicity, we also neglect the Coriolis force, ocean tilting and the atmospheric drag (we assume low-wind conditions). Under these conditions, conservation of momentum and mass lead to the following system of equations:
\begin{subequations}\label{eq:2D_conservation}
\begin{align}
	\rho H \frac{\D \bu}{\D t} &= \nabla\cdot\bsigma + \bt_o, \label{eq:2D_conservation_momentum}\\
	\frac{\D A}{\D t} &= - A\,\nabla\cdot \bu.\label{eq:2D_conservation_mass}
\end{align}
\end{subequations}
see \citet{hibler1979} or \cite{gray1994}. For any scalar or vector-valued function $f$, the material derivative is given by $\D f/\D t = \partial f/\partial t + (\bu \cdot \nabla)\,f$. Here, we assume the ice thickness $H$ to be spatially uniform for simplicity, although in general we require an additional equation, analogous to \eqref{eq:2D_conservation_mass} but in terms of $H$, for mass to be conserved. Equation \eqref{eq:2D_conservation_momentum} is a depth-averaged statement of conservation of momentum of the sea ice layer; here, $\bt_o$ is the drag force exerted by the ocean on the sea ice. Given the surface ocean velocity field $\bu_o$, this drag force is generally parameterized in terms of the drag coefficient $C_o$ and the ocean water density $\rho_o$ by 
\begin{align}\label{eq:def_to}
	\bt_o := \rho_o C_o \|\bu_o - \bu\|(\bu_o - \bu),
\end{align}
with $\|\cdot\|$ denoting the Euclidean norm of a vector. 

The conservation laws \eqref{eq:2D_conservation} must be accompanied by constitutive relations. To write these, we first decompose the Cauchy stress tensor into a pressure term $p$ and its deviatoric component $\btau$,
\begin{align}\label{eq:sigma_decomp}
	\bsigma = \btau - p\bI,
\end{align}
where $\bI$ is the identity tensor, and define the strain rate tensor $\bD$ and its deviatoric component $\bS$ as
\begin{align}
	\bD := \frac{1}{2}\left(\nabla\bu + \nabla\bu^\top\right) \quad \text{and}\quad \bS := \bD - \frac{1}{2}\left(\nabla\cdot\bu\right) \bI.
\end{align}
In the following, for a given tensor $\bT$, its second invariant is denoted by
\begin{align}
	\|\bT\|=\sqrt{\frac{1}{2}\tr{(\bT^2)}}.
\end{align} 
The fundamental idea behind the $\mu(I)$ rheology is that the constitutive relation for a granular flow depends on the inertial number, a dimensionless quantity defined as
\begin{align}\label{eq:def_I}
	I := \overline{d}\sqrt{\frac{H\rho_i}{p}}\|\bS\|,
\end{align}
where $\rho_i$ is the ice density, $\overline{d}$ denotes an average ice floe size, and $p$ is the pressure emerging in \eqref{eq:sigma_decomp}  \citep{dacruz2005, jop2006, pouliquen2006}. Throughout this document, we set $\overline{d}$ to be spatially constant over the whole domain, avoiding the need to consider the transport of this quantity. \cite{savage1984} interprets the quantity $I^2$ as the ratio between collisional (i.e.~inertial) stresses and the total shear stress exerted on the material. Accordingly, for low values of $I$, the inertial effects of grains become negligible and the flow approaches a quasi-static regime; conversely, as $I$ increases, collisional forces become increasingly important relative to the external forces exerted on the material. The functional relationship that establishes the material's rheology is written in terms of an effective friction $\mu(I)$, defined as
\begin{align}\label{eq:muI}
	\mu(I) := \frac{\|\btau\|}{p}.
\end{align}
We remark that the effective friction $\mu$ is defined in analogy with Coulomb's model of friction as the ratio between the shear (tangential) stress and the pressure (normal stress). Moreover, it is also helpful to think of the pressure $p$ as a quantification of the material's strength and its resistance to viscous and plastic deformation, as made clear in section \ref{sec:analysis}. It should be noted that the $\mu(I)$ model is a phenomenological model that has been found to work well with granular media, yet it is unclear if it represents some kind of limit for a large particle system.

To obtain a relationship between stress and strain, we need an additional constitutive law. \cite{jop2006} propose the following equality that aligns $\bS$ with $\btau$:
\begin{align}\label{eq:alignment}
	\frac{\bS}{\|\bS\|} = \frac{\btau}{\|\btau\|}.
\end{align}
Combining \eqref{eq:muI} and \eqref{eq:alignment}, the relationship between deviatoric components of the stress tensor and the shear strain can be written as 
\begin{align}\label{eq:general_stress_shear_2D}
	\btau = \mu(I)p\frac{\bS}{\|\bS\|}.
\end{align}

Compressible granular flows require a dilatancy law which relates the concentration $A$ with the inertial number $I$,
\begin{align}\label{eq:dilatancy}
	A = \Phi(I),
\end{align}
see \cite{dacruz2005}. In general, $\Phi$ is found to be a strictly decreasing function of $I$, in such a way that the concentration $A$ decreases with the rate of shearing $\|\bS\|$, a phenomenon know as dilatancy. Moreover, if $\Phi$ is strictly decreasing, it is invertible, and one can write an expression for the pressure $p$ analogous to an equation of state in thermodynamics,
\begin{align}
	p = \rho_i \overline{d}^2H\left(\frac{\|\bS\|}{\Phi^{-1}(A)}\right)^2,
\end{align}
where we have combined equations \eqref{eq:def_I} and \eqref{eq:dilatancy}. In the problems considered in this paper, we find the spatial variations in sea ice concentration to be small. Although this would make the assumption of incompressibility reasonable, the periodic one dimensional nature of these problems renders the dilatancy law \eqref{eq:dilatancy} necessary for the model to be well-posed. This point is explained below in section \ref{subsec:steady_periodic_ocean_problem}.

\subsection{The steady one dimensional periodic ocean problem}\label{subsec:steady_periodic_ocean_problem}

\begin{figure}
	\centering
	\includegraphics[scale=1]{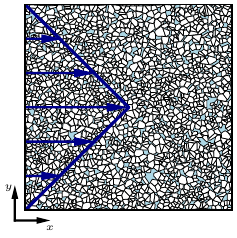}
	\caption{The periodic ocean setup. The domain is a square patch of ocean, periodic in the horizontal and vertical directions. The ice floes are driven by the ocean velocity field \eqref{eq:uo_hat} (in blue), which does not vary in the $x$-direction.}
	\label{fig:setup}
\end{figure}

The model problem considered in this paper consists of a square patch of ocean of length $L$ with periodic boundaries in both the $x$ and $y$-directions. The ice floes floating on this patch are driven by an ocean velocity field that only varies in the $y$-direction, as depicted in figure \ref{fig:setup}. We neglect time-dependent effects and only consider steady conditions in the forcing i.e.~$\bu_o(x,y, t) = (u_o(y),0)$.

This configuration renders the continuum problem one dimensional and independent of time, such that $\bu(x,y, t) = (u(y), 0)$, $A(x,y, t) = A(y)$, and $p(x,y,t) = p(y)$. In this setting, the equations for conservation of momentum \eqref{eq:2D_conservation_momentum}, together with the constitutive equation \eqref{eq:general_stress_shear_2D}, simplify to the following system on $(0,L)$:
\begin{subequations}\label{eq:one_dim_system}
\begin{align}
	- \frac{\dd\sigma_{xy}}{\dd y}= \rho_o C_o |u_o - u|(u_o - u)&\\
	\frac{\dd p}{\dd y} = 0, &\label{eq:one_dim_system_mom_y}\\
	\sigma_{xy} = \mu(I)p  \frac{\dd u/\dd y}{\left|\dd u/\dd y\right|}, &\\
	I = \overline{d}\sqrt{\frac{H\rho_i}{p}}\left|\dd u/\dd y\right|. &
\end{align}
\end{subequations}
Due to \eqref{eq:one_dim_system_mom_y}, which represents the balance of momentum in the $y$-direction, the pressure is a constant (but unknown) over the domain. \cite{dacruz2005} and \cite{herman2022} find $p$ by enforcing normal stress boundary conditions along a boundary of the domain, but we cannot do the same because the domain is periodic. In the DEM computations, which we introduce in section \ref{sec:inferring}, we set a global ice concentration $A_0\in [0,1]$ which equals the domain averaged value of the local concentration $A$, such that
\begin{align}\label{eq:conservation_mass_constraint}
	\frac{1}{L}\int_0^L A\,\dd y = A_0.
\end{align}
Therefore, if we assume the sea ice to behave like a compressible fluid, condition \eqref{eq:conservation_mass_constraint} and the dilatancy law \eqref{eq:dilatancy} close the system of equations. In sections \ref{sec:inferring} and \ref{sec:comparison_DEM}, we justify this modeling choice by demonstrating that dilatancy emerges in the DEM computations and that our model is capable of capturing it accurately. 

In this paper we only consider the following ocean velocity profile for simplicity:
\begin{align}\label{eq:uo_hat}
	u_o(y) = u_{o,max}\left(1 - |1 - 2y/L|\right),
\end{align}
for some maximum velocity $u_{o,max} > 0$. Figure \ref{fig:setup} contains a plot of this velocity profile.

\section{Inferring the constitutive equations of the system from the DEM}\label{sec:inferring}

\begin{table}
  \begin{center}
\def~{\hphantom{0}}
	\begin{tabular}{cccccc}
		$C_o$ & $\rho_i$ & $\rho_o$ & $E$ & $\nu$ & $\mu^\ast$ \\
	\midrule
		$3\times 10^{-3}$ & $900\,\si{kg}/\si{m}^3$ & $1026\,\si{kg}/\si{m}^3$& $6\times 10^6\,\si{Pa}$ & 0.3 & 0.2\\
	\end{tabular}	
  \caption{Values for material parameters used throughout this document. Here, $C_o$ is the drag coefficient for the ocean current and $\rho_i$ and $\rho_o$ are the ice and ocean water densities, respectively. The Young's modulus $E$, Poisson's ratio $\nu$, and inter-floe friction coefficient $\mu^\ast$ are used in the calculation of collisional forces, as described in Appendix \ref{app:collision}.}
  \label{tab:material_parameters}
  \end{center}
\end{table}

The system of equations presented in section \ref{subsec:steady_periodic_ocean_problem} is incomplete because we need additional expressions for $\mu(I)$ and $\Phi(I)$. In this paper, we infer these additional equations from data generated with SubZero, a DEM developed by \cite{manucharyan2022} and used for modeling sea ice dynamics with polygonal-shaped ice floes. Following the setup presented in section \ref{subsec:steady_periodic_ocean_problem}, we perform runs with $n = 2000$ floes over a square patch of ocean of length $L = 100\,\si{km}$, driven by the ocean velocity field \eqref{eq:uo_hat}. This means that the $\mu(I)$ and $\Phi(I)$ functions are inferred from DEM solutions with a constant number of floes (and therefore constant average floe size). In theory, the effects of floe size are included in the non-dimensional parameter $I$. In practice, the effects of $n$ are subtle and not too well captured with our continuum model, see figure \ref{fig:comparison_n} in section \ref{sec:comparison_DEM} below.

\begin{figure}
	\centering
	\includegraphics[scale=1]{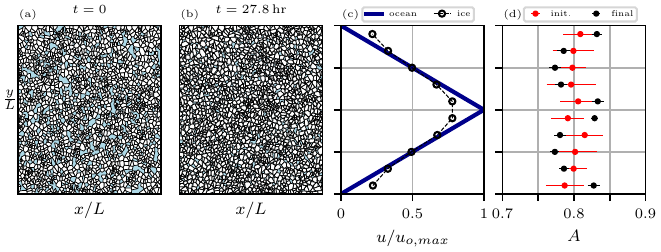}
	\caption{Emergence of dilatancy in the DEM. Here, $A_0 = 0.8$ and $u_o = 1\,\si{m}/\si{s}$. Ten runs to steady state are performed from randomized initial conditions, with (a) and (b) being examples of initial and final states, respectively. (c) ocean and sea ice velocities, (d) initial (red) and final (black) mean local sea ice concentration with the standard deviation (bars). Clearly, $A$ decreases in the areas of maximum shear strain, compare (c) and (d).}
	\label{fig:dilatancy}
\end{figure}

\begin{figure}
	\centering
	\includegraphics[scale=1]{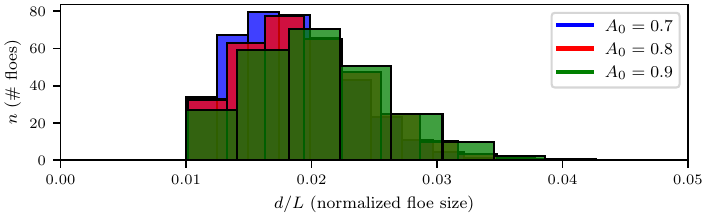}
	\caption{Floe size distributions for different global sea ice concentrations $A_0$ and a total number of ice floes $n = 2000$. The floe size $d$ is defined as the square root of the floe area.}
	\label{fig:floe_size_dist}
\end{figure}

For a given number of floes and a global sea ice concentration $A_0$, the initial configuration of ice floes is generated with SubZero's packing algorithm, which is based on a Voronoi tessellation of the domain \citep{manucharyan2022} (see panel (a) of figure \ref{fig:dilatancy} below for an example of the outcome of this packing algorithm). Defining the floe size $d$ as the square root of the floe area, this generates a polydisperse floe size distribution whose histogram we can see in figure \ref{fig:floe_size_dist} for three values of $A_0$; we find that $d/L$ is approximately between 0.01 and 0.04. Although we do not study the effects of polydispersity on the rheology here, we remark that \citet{herman2022} finds the dilatancy law $\Phi$ to vary visibly with the degree of polydispersity, while the effective friction $\mu$ presents much smaller variations. 

We run the DEM simulations for $2\times 10^4$ time steps, each of $5\,\si{s}$ (the total running time is approximately equal to 27.8 hours). In general, we find that the velocity, stress, and sea ice concentration, averaged over the last 25\% of the time steps, remain relatively unchanged when a longer computation is performed, and hence we consider that a steady state has been reached. This temporal averaging is performed over data which, at each time step, has been averaged spatially over a grid as described in appendix \ref{app:averaging}. For these simulations, the material parameters which determine the effects of the ocean drag and the collisions among floes are presented in table \ref{tab:material_parameters}. Collisional forces and the resultant stresses, which determine the fields $\bsigma$ and $p$, are computed as explained in appendix \ref{app:collision}.

In figure \ref{fig:fit} we plot the values of the friction $\mu = |\sigma_{xy}|/p$ and local concentration $A$ against $I$ for different global concentrations $A_0$ between $0.7$ and $0.95$ and different maximum ocean velocities $u_{o,max}$ between $0.1$ and $1\,\si{m/s}$. The mean ocean velocities in the ocean patch are therefore between $0.05$ and 0.5 m/s, values that are consistent with real observations \citep{stewart2019}. In all of these computations, we set the ice thickness to $H = 2\,\si{m}$. We find an increase in the friction $\mu$ and a decrease in the local concentration $A$ as $I$ increases. The decrease in the local concentration of sea ice $A$ with an increase in $I$ is due to dilatancy. Figure \ref{fig:dilatancy} presents an example of how dilatancy emerges in the DEM computations: given a random initial distribution of ice floes, when a steady state is reached the concentration $A$ decreases in the areas where the largest shearing occurs ($y = 1/4$ and $y = 3/4$), and increases elsewhere. Since the global sea ice concentration $A_0$ is constant, in this context dilatancy represents a reorganization of the local concentration profile $A(y)$.

The trends found in the data in figure \ref{fig:fit} are well fitted with the following family of functions:
\begin{align}\label{eq:fitting_functions}
	\mu(I) = \mu_0 + \mu_1 I \quad \text{and} \quad \Phi(I) = 1 - \phi_0 I^\alpha.
\end{align}
A linear behavior is also found for $\mu$ in \cite{dacruz2005}. The four parameters $(\mu_0,\mu_1,\phi_0,\alpha)$ are calculated by minimizing the least-squares misfit problem between the points in figure \ref{fig:fit} and the functions in \eqref{eq:fitting_functions}. The resulting values are shown in table \ref{tab:fitting_parameters}. For the remainder of the document, any numerical solution of the one dimensional system \eqref{eq:one_dim_system} is solved by setting the parameters in functions \eqref{eq:fitting_functions} to the values given in table \ref{tab:fitting_parameters}.

\begin{table}
  \begin{center}
\def~{\hphantom{0}}
	\begin{tabular}{cccc}
		$\mu_0$ & $\mu_1$ & $\phi_0$ & $\alpha$ \\
	\midrule
		0.26    & 4.93   & 0.53     & 0.24\\
	\end{tabular}
  \caption{Parameters for the functions $\mu(I)$ and $\Phi(I)$ in \eqref{eq:fitting_functions} obtained by minimizing the least-squares misfit with the data plotted in figure \ref{fig:fit}. These are the numerical values used for computing solutions to the continuum model \eqref{eq:one_dim_system} throughout this paper.}
  \label{tab:fitting_parameters}
  \end{center}
\end{table}

Departure from the fitting curves are most visible when the ocean velocities and sea ice concentrations are low, see the case where $u_{o,max} = 0.1\,\si{m}/\si{s}$ and $A_0 = 0.7$. Unsurprisingly, in section \ref{sec:comparison_DEM} below, we also find the greatest misfit between the DEM and the continuum model precisely in this setting, when $u_{o,max} = 0.1\,\si{m}/\si{s}$ and $A_0 = 0.7$, see panel (m) in figure \ref{fig:comparison_fitted} below. In particular, in this setting, the fundamental balance between shear stress and ocean drag in the DEM is found to no longer hold, see section \ref{sec:comparison_DEM}. 

The constitutive equation in 2D resulting from functions \eqref{eq:fitting_functions} is the sum of a plastic and a viscous term:
\begin{align}\label{eq:rheology_2D}
	\btau = \underbrace{\mu_0 p \frac{\bS}{\|\bS\|}}_\text{plastic} + \underbrace{\mu_1  \overline{d}^2 \sqrt{\rho_i H p}\, \bS}_\text{viscous}.
\end{align}
A consequence of this linear behavior is that $\mu$ is approximately constant for small values of $I$, and this is precisely what we see for $I < 10^{-2}$ in figure \ref{fig:fit}. For $I\ll 1$, we have that $\mu(I) \approx \mu_0$, and therefore it is the plastic term that dominates the rheology. This is essentially the quasi-static regime, where collisions are negligible. This plastic term follows from a Mohr-Coulomb yield criterion with an internal angle of friction $\tan^{-1}(\mu_0)$; examples of the Mohr-Coulomb yield criterion used for sea ice modeling can be found in \cite{ip1991}, \cite{gutfraind1997} and \cite{ringeisen2019}. The viscous term, which becomes increasingly important as the inertial number $I$ increases, can be associated with the collisional component of the rheology. A viscous rheology is derived in \cite{shen1987} for modeling the rheological effects of collisions in sea ice, which, as we explain in section \ref{sec:similarities_model}, is very similar to the viscous component in \eqref{eq:rheology_2D}.

\begin{figure}
	\centering
	\includegraphics[scale=1]{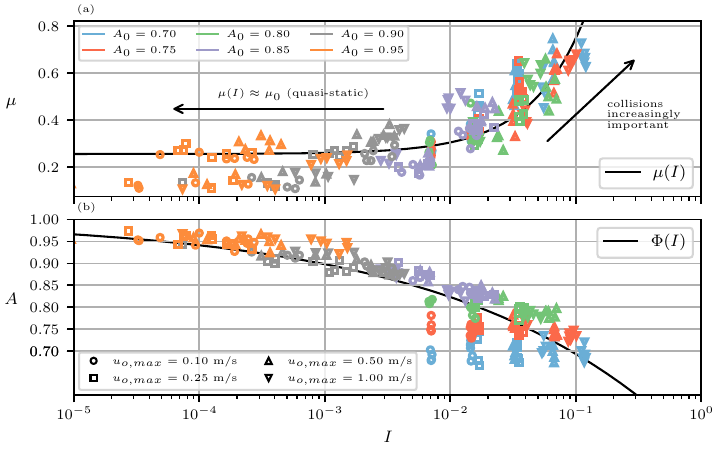}
	\caption{(a) friction $\mu = |\sigma_{xy}|/p$ and (b) concentration $A$ against $I$. Ten friction and concentration values are extracted from each DEM run by averaging along the grid cells plotted in figure \ref{fig:inference}. The black lines correspond to the functions \eqref{eq:fitting_functions} fitted to the data by minimizing the least-squares misfit.}
	\label{fig:fit}
\end{figure}

\section{Analysis of the inferred continuum model}\label{sec:analysis}

This section focuses on the one dimensional system of equations presented in section \ref{subsec:steady_periodic_ocean_problem}, with the functions $\mu$ and $\Phi$ taking the form \eqref{eq:fitting_functions}. Firstly, in section \ref{subsec:nondim} we non-dimensionalize the system of equations to understand the relative importance of the different terms involved. 

The remaining two sections explore the existence and uniqueness of solutions to this system, a regularization technique that facilitates its numerical solution, and the behavior of solutions under different limits. In particular, in section \ref{subsec:momentum_eq}, we focus only on the momentum equation \eqref{eq:one_dim_system_mom_y}. In section \ref{subsubsec:minimization}, we show that the momentum equation can be rewritten as a minimization problem. This allows us to establish that solutions to this equation exist and are unique, and it allows us to make sense of the plastic component (see \eqref{eq:rheology_2D}) in a rigorous sense (as a variational inequality). Moreover, this equivalence also motivates a regularization of the plastic term that simplifies its numerical solution considerably, as described in section \ref{subsubsec:regularization}. We end section \ref{subsec:momentum_eq} with an analysis of how the sea-ice velocity behaves under different limits in parameter values: in section \ref{subsubsec:limit_behaviour_pressure}, we explore the behavior of the sea-ice velocity for small and large pressures $p$; then, in section \ref{subsubsec:plastic}, we derive an analytical solution for the sea-ice velocity in the purely plastic limit. Understanding these limit solutions for the velocity is helpful for interpreting the DEM results we present in section \ref{sec:comparison_DEM}. It is also useful for the analysis we present in section \ref{subsec:complete_model}, which considers the complete system of equations \eqref{eq:one_dim_system}. We begin by presenting a numerical method for solving the complete system in section \ref{subsubsec:numerics}. Then, in section \ref{subsubsec:existence_uniqueness_complete_system}, we sketch out a demonstration of the existence and uniqueness of solutions to the complete system. 

We remark that sections \ref{subsubsec:minimization}, \ref{subsubsec:regularization}, and \ref{subsec:complete_model} are mostly concerned with questions of a mathematical and numerical nature. Although we believe these to be important topics in establishing the suitability of our model for modeling purposes, they are not required for understanding the remainder of the paper. We also note that, for simplicity, the solutions presented throughout this section result from driving the ice floes with the ocean velocity profile \eqref{eq:uo_hat}, although the analysis can be extended to more general ocean velocity profiles by following the same steps. 

\subsection{Non-dimensionalization of the system}\label{subsec:nondim}

For the non-dimensionalization, we set the characteristic magnitudes 
\begin{align}
	[y] = L,\quad [H] = H,\quad [u] = u_{o,max}, \quad \text{and} \quad [\sigma] = \rho_i [u]^2 [H]
\end{align}
for the length, thickness, velocity, and stress, respectively. We scale the velocities $u$ and $u_o$ with $[u]$, the spatial variables $y$ and $\overline{d}$ with $[y]$, the thickness $H$ with $[H]$, and $\sigma_{xy}$ and $p$ with $[\sigma]$. 

From this point onward, all variables considered are non-dimensional unless the contrary is made explicitly clear or units are specified. Keeping the same notation as used for dimensional variables, the following normalized system of equations is derived for $u$, $I$, and $A$, and for the constant $p > 0$:
\begin{subequations}\label{eq:one_dim_system_normal}
\begin{align}
	- \epsilon\frac{\dd}{\dd y} \left( \mu_0 p \frac{\dd u/\dd y}{|\dd u/\dd y|} + \mu_1 \sqrt{p\frac{A_0}{n}}\frac{\dd u}{\dd y} \right)= \beta_o |u_o - u|(u_o - u), \label{eq:mom_normal}&\\
	I = \sqrt{\frac{A_0}{pn}}\left|\dd u/\dd y\right|. \label{eq:I_normal}&\\
	A = 1 - \phi_0 I^\alpha, \label{eq:A_normal}&\\
	\int_0^1A\,\dd y = A_0.\label{eq:A_constraint_normal}
\end{align}
\end{subequations}
Here, $\epsilon = H/L$ and $\beta_o = \rho_o/\rho_iC_o$; the non-dimensional average floe size $\overline{d}$ is set to $\sqrt{A_0/n}$. The system \eqref{eq:one_dim_system_normal} is closed by enforcing periodic boundary conditions for $u$, $I$, and $A$. Following our findings in table \ref{tab:fitting_parameters}, we assume that the parameters $\mu_0$, $\phi_0$, and $\alpha$ are strictly greater than zero. We also assume $\mu_1 > 0$ in all but section \ref{subsubsec:plastic}, where we study the case when $\mu_1 = 0$ with the intention of understanding the plastic component of the momentum equation \eqref{eq:mom_normal}.

All numerical results computed in this section take \eqref{eq:uo_hat} as the ocean velocity, which is written as
\begin{align}\label{eq:uo_hat_normal}
	u_o(y) = 1 - |1 - 2y|
\end{align}
for $y\in(0,1)$ when non-dimensionalized.

\subsection{The momentum equation}\label{subsec:momentum_eq}

In order to understand the system of equations \eqref{eq:one_dim_system_normal}, we first focus on the momentum equation \eqref{eq:mom_normal}. When considering the entire system \eqref{eq:one_dim_system_normal}, the pressure $p\in\RR$ is one of the unknowns. However, it is useful to first assume it to be known, in which case we can solve the momentum equation \eqref{eq:mom_normal} for $u$ and study the effect of $p$ on $u$. Here, we show that \eqref{eq:mom_normal} can be understood as a minimization problem. This reformulation of the momentum equation allows us to establish the existence and uniqueness of solutions. Moreover, the optimality conditions for the minimization problem result in a different formulation of the plasticity component of the rheology which avoids the singularity, present in \eqref{eq:mom_normal}, at $\dd u/\dd y = 0$. With this reformulation of the plastic term, we are able to find analytical solutions to the purely plastic problem which arises when $I \ll 1$, near the quasi-static regime.

\subsubsection{Reformulation of \eqref{eq:mom_normal} as a minimization problem}\label{subsubsec:minimization}

Given a pressure $p > 0$, solutions $u$ to \eqref{eq:mom_normal} minimize the following functional,
\begin{equation}\label{eq:Jc}
	\Jc(u) := \epsilon\mu_0p \int_0^1\left|\frac{\dd u}{\dd y}\right|\,\dd y + \frac{\epsilon\mu_1}{2} \sqrt{p\frac{A_0}{n}} \int_0^1\left|\frac{\dd u}{\dd y}\right|^2\,\dd y + \frac{\beta_o}{3}\int_0^1|u_o - u|^3\,\dd y
\end{equation}
over the space of velocity profiles
\begin{align}\label{eq:def_V}
	V := \left\lbrace v\in H^1((0,1)) : \text{$v$ is a periodic function}\right\rbrace.
\end{align}
In the definition of $V$, the space $H^1((0,1))$ denotes the Sobolev space of weakly differentiable and periodic functions on the unit interval \citep{adams2003}. As explained in appendix \ref{app:existence}, the functional $\Jc$ is strictly convex and coercive over $V$ and therefore admits a unique minimizer. In this sense, one can state that the momentum equation \eqref{eq:mom_normal} also has a unique solution.

To derive \eqref{eq:mom_normal} for the minimizer $u$ of $\Jc$, we first note that, if $u$ minimizes $\Jc$, then
\begin{align}
	\frac{1}{t}\left(\Jc(u + t(v-u)) - \Jc(u)\right) \geq 0 \quad \forall v\in V,\, \forall t\in (0,1).
\end{align}
The three terms in the right hand size of \eqref{eq:Jc} are convex, with the two last ones differentiable over all $V$. By exploiting the convexity of the first term (the $L^1$ norm) and the differentiability of the other two terms, we find that
\begin{align}\label{eq:vi}
	\begin{split}
	\epsilon\mu_0p \left( \left|\frac{\dd v}{\dd y}\right| - \left|\frac{\dd u}{\dd y}\right| \right) + \epsilon\mu_1\sqrt{p\frac{A_0}{n}} \int_0^1 \frac{\dd u}{\dd y}\frac{\dd (v-u)}{\dd y}\,\dd x \\
	- \beta_o \int_0^1 |u_o - u|(u_o - u)(v-u)\,\dd x \geq 0 \quad \forall v\in V.
	\end{split}
\end{align}
A variational statement as in \eqref{eq:vi} is known as a variational inequality. Under the assumption that the solution $u$ is not only in $V$ but is twice continuously differentiable, we may deduce that
\begin{subequations}\label{eq:mom_no_sing}
	\begin{align}
		- \epsilon\frac{\dd }{\dd y}\left( \sigma_{xy}^P + \mu_1 \sqrt{p\frac{A_0}{n}}\frac{\dd u}{\dd y} \right)= \beta_o |u_o - u|(u_o - u), \label{eq:mom_no_sing_eq}\\
		|\sigma_{xy}^P| \leq \mu_0 p, \label{eq:plast-bound}\\
		\sigma_{xy}^P  \left|\frac{\dd u}{\dd y}\right| = \mu_0 p\frac{\dd u}{\dd y} \label{eq:plast-constitutive}.
	\end{align}
\end{subequations}
A similar derivation to that of \eqref{eq:mom_no_sing} from \eqref{eq:vi} can be found in \cite[section 1.3]{glowinski1981}. In \eqref{eq:mom_no_sing}, we have introduced $\sigma_{xy}^P$, the purely plastic component of the shear stress. Introducing this new variable allows us to reformulate \eqref{eq:mom_normal} such that the singularity at $\dd u/\dd y = 0$ is removed. Indeed, if $\dd u/\dd y \neq 0$, it is clear that \eqref{eq:mom_no_sing} is equivalent to \eqref{eq:mom_normal}. In this case, we have that $|\sigma_{xy}^P| = \mu_0 p$ and we say that the material has reached its plastic yield strength $\mu_0 p$. Conversely, when $\dd u/\dd y = 0$, \eqref{eq:mom_no_sing_eq} remains well defined, unlike \eqref{eq:mom_normal}. We remark that $\dd u/\dd y = 0$ must follow from \eqref{eq:plast-constitutive} whenever $|\sigma_{xy}^P| < \mu_0 p$ (the material has not reached its plastic yield strength). Below, in section \ref{subsubsec:plastic}, we provide further insight into the plastic component of the shear stress by computing purely plastic solutions to the momentum equation analytically.

\subsubsection{Regularization of the plastic term to facilitate its numerical solution}\label{subsubsec:regularization}

\begin{figure}
	\centering
	\includegraphics[scale=1]{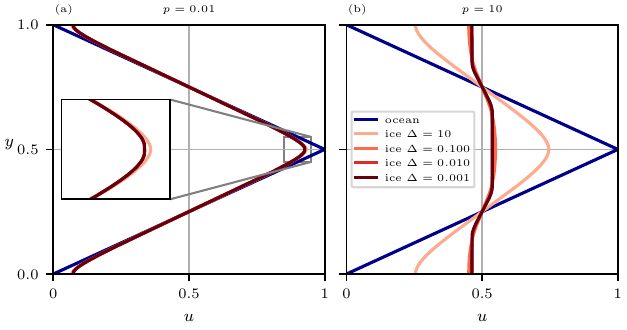}
	\caption{Solutions $u$ to the momentum equation \eqref{eq:mom_reg} for different values of the ice pressure $p$ and the numerical regularization parameter $\Delta$. We set $u_o$ equal to \eqref{eq:uo_hat}, normalized with $[u] = u_{o,max}$, and $\epsilon = 2\times 10^5$, $A_0 = 0.8$ and $n = 2000$. We observe convergence to a solution as $\Delta \to 0$ and, for small $p$, $u \to u_o$ in (a) and, for large $p$, $u\to 0.5$ in (b), as expected from the theory.}
	\label{fig:momentum_sol_diff_p}
\end{figure}

The first order optimality condition for the minimization of $\Jc$ is a variational inequality (rather than a variational equality) because the first term to the right hand side of \eqref{eq:Jc} (the $L^1$ norm) is non differentiable when $\dd u/\dd y = 0$. We can make $\Jc$ differentiable by regularizing it as follows:
\begin{equation}\label{eq:Jc_smooth}
	\Jc_\Delta(u) := \epsilon\mu_0p \int_0^1\sqrt{\left|\frac{\dd u}{\dd y}\right|^2 + \Delta^2}\,\dd y + \frac{\epsilon\mu_1}{2} \sqrt{p\frac{A_0}{n}} \int_0^1\left|\frac{\dd u}{\dd y}\right|^2\,\dd y + \frac{\beta_o}{3}\int_0^1|u_o - u|^3\,\dd y,
\end{equation}
where $\Delta > 0$ is a small parameter. The first order optimality conditions for the minimization of $\Jc_\Delta$ over $V$ corresponds with the following equation:
\begin{align}\label{eq:mom_reg}
	- \epsilon\frac{\dd }{\dd y}\left( \mu_0 p \frac{\dd u/\dd y}{\sqrt{|\dd u/\dd y|^2 + \Delta^2}} + \mu_1 \sqrt{p\frac{A_0}{n}}\frac{\dd u}{\dd y} \right)= \beta_o |u_o - u|(u_o - u).
\end{align}
Although the system \eqref{eq:mom_no_sing} can be solved numerically by e.g.~introducing a Lagrange multiplier \citep{glowinski1981}, it is easier to solve equation \eqref{eq:mom_reg}. This is the strategy we take for solving the momentum equation and, as we explain below in section \ref{subsubsec:numerics}, the complete system \eqref{eq:one_dim_system_normal}. To do so, we use the finite element method (FEM) implemented in Firedrake \citep{firedrake}. In particular, we approximate the velocity profile $u$ with continuous piece-wise linear functions. In figure \ref{fig:momentum_sol_diff_p}, we plot solutions to \eqref{eq:mom_reg} for two different values of $p$ and a range of $\Delta > 0$. Convergence of the velocity profiles as $\Delta \to 0$ is clearly visible in these figures; in fact, for $\Delta \leq 10^{-2}$, the solutions become indistinguishable. We remark that, if we remove the viscous component of the rheology in the regularized equation \eqref{eq:mom_reg}, we essentially arrive at Hibler's model in one dimension, given below by \eqref{eq:hibler-mom}.

\subsubsection{Velocity profiles in the limit of small and large pressures}\label{subsubsec:limit_behaviour_pressure}

The pressure or ice strength $p$ is a fundamental variable in the continuum model; understanding its effect on $u$ is fundamental for making sense of our sea ice model. Figure \ref{fig:momentum_sol_diff_p} suggests that, for small $p$, the velocity profile $u$ approaches the ocean's $u_o$ and, for large $p$, $u$ flattens and comes close to a constant-valued velocity profile. We can deduce this behavior from the functional $\Jc$. For small values of $p$,
\begin{align}\label{eq:Jc_lim_small_p}
	\lim_{p\to 0} \Jc(u) = \frac{\beta_o}{3}\int_0^1|u_o - u|^3\,\dd y,
\end{align}
and therefore, since $u$ minimizes $\Jc$, it must follow that $u \to u_o$. On the other hand, for large values of $p$, we see that 
\begin{align}\label{eq:Jc_lim_large_p}
	\Jc(u) \approx \epsilon\mu_0p \int_0^1\left|\frac{\dd u}{\dd y}\right|\,\dd y \quad \text{as $p\to \infty$}
\end{align}
and, in principle, any constant velocity profile $u_c \in \RR$ minimizes \eqref{eq:Jc_lim_large_p}. However, this constant velocity field is constrained by the total force balance of the system. That is, due to the periodic boundary conditions, if we integrate \eqref{eq:mom_no_sing_eq} along $(0,1)$, we must have that
\begin{align}\label{eq:momentum_integral_statement}
	\beta_o \int_0^1 |u_o - u|(u_o - u)\,\dd y  = 0.
\end{align}
Therefore, the constant value to which $u$ tends as $p\to\infty$ will be a solution to \eqref{eq:momentum_integral_statement} with $u = u_c \in \RR$. Below, in section \ref{subsubsec:plastic}, we show that a critical pressure $p_c$ can be found such that $u$ is constant whenever $p > p_c$ and $I \ll 1$.

\subsubsection{Purely plastic solutions to the momentum equation}\label{subsubsec:plastic}

In figure \ref{fig:fit} we can see that $\mu(I)$ roughly becomes constant for small inertial numbers $I \ll 1$, such that $\mu(I) \approx \mu_0$ and the flow rheology is plastic. This regime is closely related to the quasi-static regime for granular media, with the material behaving like a purely plastic flow characterized by a critical state at which plastic deformation occurs \citep{wood1990}.

The momentum equation to the purely plastic problem where $\mu(I) = \mu_0$ is given by 
\begin{align}\label{eq:purely-plastic-mom}
	\epsilon\frac{\dd \sigma_{xy}^P}{\dd y} = - \beta_o|u_o - u|(u_o - u).
\end{align}
Equation \eqref{eq:purely-plastic-mom} must be complemented with \eqref{eq:plast-bound} and \eqref{eq:plast-constitutive}. Here, we present a method for calculating solutions to \eqref{eq:purely-plastic-approximate}. Additionally, we find a critical pressure $p_c$ such that, for $p > p_c$, the velocity profiles $u$ that solve \eqref{eq:purely-plastic-mom} remain constant and no shear strain occurs in the sea ice. In section \ref{sec:comparison_DEM} below, we show that this critical pressure approximates the pressure computed from the DEM when the global sea ice concentration is high. When following the derivation of purely plastic solutions, it is helpful to look at their plots in figure \ref{fig:pure_plastic} below.

Conditions \eqref{eq:plast-bound} and \eqref{eq:plast-constitutive} for the plastic stress tensor indicate that there exist two distinct regions of the flow field: a region where the sea ice has yielded and $|\sigma_{xy}^P| = p\mu_0$, and another region where the ice has not yielded and $|\sigma_{xy}^P| \leq p\mu_0$ and $\dd u/\dd y = 0$. By working with this distinction, we can find a purely plastic solution to \eqref{eq:purely-plastic-mom}. Due to the symmetry of the problem, we assume that $\sigma_{xy}^P = 0$ at $y = 0$, $1/2$, and $1$. Then, integrating \eqref{eq:purely-plastic-mom} along the interval $(0,y)$ for some $y\in(0,1)$, we find that 
\begin{align}
	\sigma_{xy}^P(y) = -\frac{\beta_o}{\epsilon}\int_0^y|u_o - u|(u_o - u)\,\dd y.
\end{align}
Since $\sigma_{xy}^P(0) = 0$, we must necessarily have an interval $(0,y_1)$ where the ice has not yielded and the velocity equals a constant $u_1$. In the context of the ocean velocity profile \eqref{eq:uo_hat_normal}, it makes sense to assume that $u_1 > 0$, and therefore $u_1 \geq u_o$ near $y = 0$, so that
\begin{align}\label{eq:purely-plastic-stress-1}
	\sigma_{xy}^P(y) = \frac{\beta_o}{6\epsilon}\left( (2y - u_1)^3 + u_1\right).
\end{align}
Since the material has not yielded for $y\in (0,y_1)$, we have that $|\sigma_{xy}^P(y)| < \mu_0 p$. Equation \eqref{eq:purely-plastic-stress-1} tells us that $\sigma_{xy}^P(y)$ increases with $y$ over this interval; this means that  
\begin{align}
	\lim_{y\to y_1; y < y_1} \sigma_{xy}^P(y) = \mu_0 p,
\end{align}
and we find that 
\begin{align}\label{eq:u1_unbounded}
	u_1 = \left(\frac{6\epsilon\mu_0p}{\beta_0}\right)^{1/3}. 
\end{align}
Clearly, an upper bound is needed for $u_1$ in \eqref{eq:u1_unbounded}, because it grows indefinitely with $p$, yet it is senseless for the ice to circulate at speeds greater than the maximum ocean velocity when a steady state has been reached. We can make sense of this paradox by firstly assuming the existence of an interval $(y_1, y_2)$, where $y_1 < y_2 < 1/2$, in which the sea ice has yielded and $\sigma_{xy}^P = \mu_0 p$. In this interval we must have that $u= u_o$ because $\sigma_{xy}^P$ is constant and therefore the ocean drag is zero. This means that $y_1 = u_1/2$. Moreover, repeating the same argument as that used for deriving \eqref{eq:u1_unbounded}, we assume that $u = u_2$ for some constant $u_2 < 1$ on $(y_2, 1/2)$ and find that $u_2 = 1 - u_1$ and $y_2 = u_2/2$. Now, the assumption that $y_1 < y_2 < 1/2$ will only hold for values of $p$ for which $u_1 \leq u_2$; that is, $u_1 \leq 1/2$, and this upper bound on $u_1$ defines a critical pressure $p_c$ given by
\begin{align}\label{eq:pcritical}
	p_c = \frac{\beta_o}{48\epsilon\mu_0}.
\end{align}
For $p \geq p_c$, the integral force balance along the domain must hold, see \eqref{eq:momentum_integral_statement}; as a result, $u_1$ can be at most equal to $1/2$. Putting these results together, we may write the analytical solution to \eqref{eq:purely-plastic-mom} as
\begin{subequations}\label{eq:purely-plastic-approximate}
\begin{align}
	& u(y) = \left\lbrace \begin{array}{cl}
		u_1 & \text{ $0<y<\frac{u_1}{2}$}, \\
		u_o(y) & \text{$y_1<y<\frac{1}{2} - y_1$}, \\
		1 - u_1 & \text{$\frac{1}{2} - y_1<y<\frac{1}{2}$},
	\end{array} \right. && \text{if $ p < p_c$},\\
	& u(y) = \frac{1}{2} && \text{if $ p \geq p_c$}.
\end{align}
\end{subequations}
for $y\in [0,1]$. We test the validity of \eqref{eq:purely-plastic-approximate} by showing that it is indistinct from its numerical counterpart in figure \ref{fig:pure_plastic}. We compute this numerical solution by regularizing the shear stress $\sigma_{xy}^P$ as in \eqref{eq:mom_reg} and setting $\Delta = 10^{-3}$.

\begin{figure}
	\centering
	\includegraphics[scale=1]{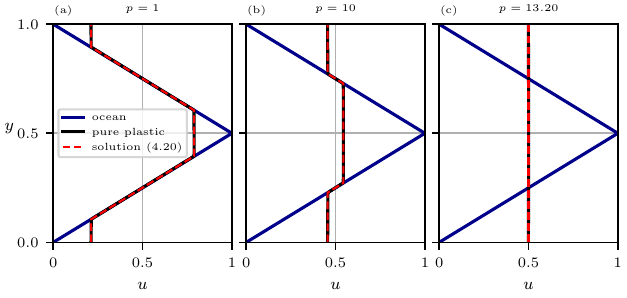}
	\caption{Solutions $u$ to the purely plastic momentum equation \eqref{eq:purely-plastic-mom}. We use the same parameter values as those used for figure \ref{fig:momentum_sol_diff_p} and $\Delta = 10^{-2}$. The numerical solutions (black) to the purely plastic problem are indistinguishable from the analytical solution given by \eqref{eq:purely-plastic-approximate} (red, dotted). As the ice pressure $p$ increases, the velocity profiles flattens at the critical pressure $p_c$ given by \eqref{eq:pcritical} and $p_c \approx 13.2$ in this case.}
	\label{fig:pure_plastic}
\end{figure}

\subsection{Solutions to the complete continuum model}\label{subsec:complete_model}

In section \ref{subsec:momentum_eq}, we have seen that, given a value for the pressure $p$, we can solve the momentum equation and find a velocity profile $u$ for the sea ice. We also prove that solutions to the momentum equation must exist and be unique. However, in general, the pressure $p$ is one of the unknowns in the system of equations \eqref{eq:one_dim_system_normal}, together with the sea ice concentration $A$ and the inertial number $I$. Here, we first present a numerical method for solving \eqref{eq:one_dim_system_normal} and show that solutions to this system always exist and, under some circumstance, are unique.

\subsubsection{A numerical method for the complete model \eqref{eq:one_dim_system_normal}}\label{subsubsec:numerics}

In order to solve the system \eqref{eq:one_dim_system_normal}, we follow the approach discussed in section \ref{subsubsec:regularization} for solving the momentum equation. There, the regularized equation \eqref{eq:mom_reg} is solved numerically with the FEM. When solving the complete system \eqref{eq:one_dim_system_normal}, we find that also regularizing the inertial number $I$ improves the convergence properties of our numerical solver substantially. Therefore, we solve for 
\begin{align}\label{eq:I_reg}
	I_\Delta := \sqrt{\frac{A_0}{pn}\left( \left|\frac{\dd u}{\dd y}\right|^2 + \Delta^2\right)} \quad \text{and} \quad A = 1 - \phi_0I_\Delta^\alpha
\end{align}
instead of \eqref{eq:I_normal} and \eqref{eq:A_normal}. Then, we use the FEM to solve the system of equations given by the regularized momentum equation \eqref{eq:mom_reg}, \eqref{eq:I_reg}, and the constraint for global concentration \eqref{eq:A_constraint_normal}. We approximate $u$ with continuous piece-wise linear functions and $I_\Delta$ and $A$ with piece-wise constant functions. Our solver is implemented in Firedrake \citep{firedrake} in such a way that the complete nonlinear system is solved with Newton's method. For small values of $\Delta$, Newton's method tends to fail unless a very good initial guess for the solution $(u,I_\Delta,A,p)$ is given. For this reason, we find that solving for a sequence of decreasing values of $\Delta$, using the solution for the previous $\Delta$ as the initial guess for the next $\Delta$, yields a robust solution method.

\begin{figure}
	\centering
	\includegraphics[scale=1]{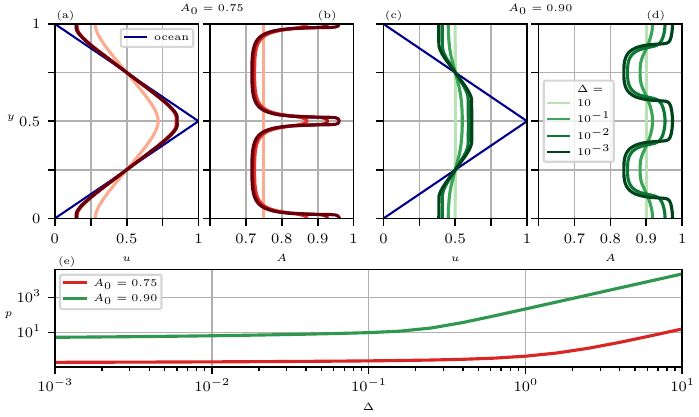}
	\caption{Solutions $u$ and $A$ to the continuum model \eqref{eq:one_dim_system_normal} with $A_0 = 0.75$ ((a),(b)) and $A_0 = 0.9$ ((c),(d)), regularized with a parameter $\Delta$ as explained in section \ref{subsubsec:numerics}. In (e) we plot the variation of the pressure $p$ with $\Delta$. We use the same parameter values as those used for figure \ref{fig:momentum_sol_diff_p}.}
	\label{fig:solutions_compete_system}
\end{figure}

We test the sensitivity of solutions $(u,I_\Delta,A,p)$ to changes in $\Delta$ by solving the system for values of $\Delta$ between $10^{-3}$ and $10$ and for $A_0 = 0.75$ and $A_0 = 0.9$. The numerical results, which are plotted in figure \ref{fig:solutions_compete_system}, indicate that the solutions become more sensitive to $\Delta$ as $A_0$ increases; this is natural, since $I$ decreases with $A_0$ and the plasticity term becomes more important. It is also clear from these plots that, although the velocity profiles $u$ come very close to convergence for the smallest values of $\Delta$, the local concentration profiles still experience visible changes around the symmetry points $y = 0$, $0.5$, and 1. In these points the shear strain drops to 0 and, by the definition of $I$, we expect $A = 1$ there. However, as we argue below in section \ref{sec:comparison_DEM}, when comparing the continuum model and the DEM, we consider such drastic changes in the local concentration to be artificial. This argument motivates the use of $\Delta$ not just as a numerical parameter that helps us solve the system numerically, but as a parameter that improves the model and may have a physical significance.

\subsubsection{Existence and uniqueness of solutions}\label{subsubsec:existence_uniqueness_complete_system}

We conclude the analysis of the continuum model by showing that at least one solution to \eqref{eq:one_dim_system} must exist and, whenever $u_o$ is given by \eqref{eq:uo_hat}, is unique. To do so, we first reformulate \eqref{eq:one_dim_system} solely in terms of $u$ and $p$ by eliminating $A$ and $I$ as follows. Substituting \eqref{eq:A_normal} into \eqref{eq:A_constraint_normal} yields
\begin{align}
	\int_0^1 I^\alpha\,\dd y = \frac{1 - A_0}{\phi_0}.
\end{align} 
Then, by substituting the definition of $I$, equation \eqref{eq:I_normal}, into the integrand in the expression above, we arrive at
\begin{align}\label{eq:constraint_in_terms_of_u}
	\int_0^1\left| \frac{\dd u}{\dd y}\right|^\alpha\dd y = \frac{1 - A_0}{\phi_0}\left(p\frac{n}{A_0}\right)^{\frac{\alpha}{2}}.
\end{align}
Therefore, the system of equations \eqref{eq:one_dim_system_normal} is equivalent to solving the momentum equation \eqref{eq:mom_normal} together with the constraint \eqref{eq:constraint_in_terms_of_u} for $u$ and $p$. We can interpret this problem as the minimization of the functional \eqref{eq:Jc} over a set of velocity profiles subject to the constraint \eqref{eq:constraint_in_terms_of_u}. Next, we define $\Fc:\RR_+\to\RR_+$ by
\begin{align}\label{eq:F}
	\Fc(p) := \int_0^1\left| \frac{\dd u}{\dd y}\right|^\alpha\dd y,
\end{align}
with $u$ denoting the solution to \eqref{eq:mom_normal} with the pressure set to $p$; this operator is well-defined because, given a pressure $p > 0$, a unique velocity profile $u$ solves the momentum equation \eqref{eq:mom_normal}. We also define $\Cc:\RR_+\to\RR_+$, given by the left hand side of \eqref{eq:constraint_in_terms_of_u}, that is, 
\begin{align}\label{eq:C}
	\Cc(p) := \frac{1 - A_0}{\phi_0}\left(p\frac{n}{A_0}\right)^{\frac{\alpha}{2}}.
\end{align}
It is then clear that a pressure $p$ is part of the solution to the system of equations \eqref{eq:one_dim_system_normal} if and only if 
\begin{align}\label{eq:neccessary_conditions_for_existence}
	\Cc(p) =\Fc(p).
\end{align}
In section \ref{subsubsec:limit_behaviour_pressure} we show that the velocity profile $u$ has two distinct limit behaviors. We find that $u$ approaches $u_o$ as $p \to 0$ and that $\mathrm{d}u/\dd y$ tends to $0$ as $p\to\infty$. This means that
\begin{align}
	\lim_{p\to 0} \Fc(p) = \int_0^1\left| \frac{\dd u_o}{\dd y}\right|^\alpha\dd y \quad \text{and} \quad \lim_{p\to \infty} \Fc(p) = 0.
\end{align}
On the other hand, the function $\Cc$ is strictly increasing, with $\Cc(0) = 0$. Therefore, whenever $\int_0^1\left| \dd u_o/\dd y\right|^\alpha\dd y > 0$ (that is, $u_o$ is not a flat velocity profile) and $\Fc$ is a continuous function, we must have at least one solution $p$ to \eqref{eq:neccessary_conditions_for_existence}. Moreover, if $\Fc$ is a decreasing function, this pressure must be unique.

\begin{figure}
	\centering
	\includegraphics[scale=1]{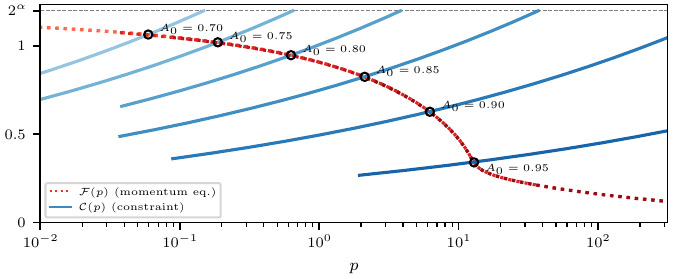}
	\caption{Functions $\Fc(p)$ and $\Cc(p)$, defined in \eqref{eq:F} and \eqref{eq:C}, respectively, for different values of $A_0$. The lighter tones of blue and red are associated with lower values of $A_0$, although $\Fc$ hardly changes with $A_0$. A pressure $p$ solves \eqref{eq:one_dim_system} whenever $\Cc(p) =\Fc(p)$ (circles). To generate  these curves numerically, the same problem setting as that of figure \ref{fig:momentum_sol_diff_p} is used.}
	\label{fig:existence}
\end{figure}

In figure \ref{fig:existence}, we plot the functions $\Cc$ and $\Fc$ for several values of $A_0$ and an ocean velocity profile given by \eqref{eq:uo_hat_normal}. In this case, we have that
\begin{align}\label{eq:int_uo}
	\int_0^1\left| \frac{\dd u_o}{\dd y}\right|^\alpha\dd y = 2^\alpha
\end{align}
and the function $\Fc$ appears to be strictly decreasing. This is expected, because an increase in pressure is accompanied by a flattening of the sea ice velocity profile. This means that there is a unique $p$ for which \eqref{eq:neccessary_conditions_for_existence} holds; since there is only one velocity profile $u$ that solves the momentum equation \eqref{eq:mom_normal}, it follows that the solution to the continuum model \eqref{eq:one_dim_system_normal} must be unique.

\section{Comparison of the DEM with the continuum model}\label{sec:comparison_DEM}

The continuum model \eqref{eq:one_dim_system_normal} is designed with the objective of capturing the averaged behavior of the sea ice simulations carried out with the DEM. Here, we verify that, in the one dimensional setting of the steady ocean periodic problem, the continuum model is indeed capable of replicating most of the results of the DEM. We remind the reader that the DEM solutions can be considered steady and one-dimensional only in the sense that, for spatially and temporally-averaged data, we can expect variations in time and in the $x$-direction to be negligible. At a small scale, we certainly expect the data to present variations in time and in the $x$-direction, and the vertical velocity of the sea-ice to be non-zero. Moreover, DEM computations are initialized using a random floe initialization; however, as displayed in figure \ref{fig:dilatancy}, the (spatially and temporally-averaged) steady states that the DEM computations reach from different floe initializations are almost indistinguishable.

Throughout this section, we use the parameters in tables \ref{tab:material_parameters} and \ref{tab:fitting_parameters} when solving the continuum model. The continuum model is solved with the FEM as explained in section \ref{subsubsec:numerics}, setting $\Delta = 10^{-3}$ and using a uniform mesh of 300 cells. 
 
\begin{figure}
	\centering
	\includegraphics[scale=1]{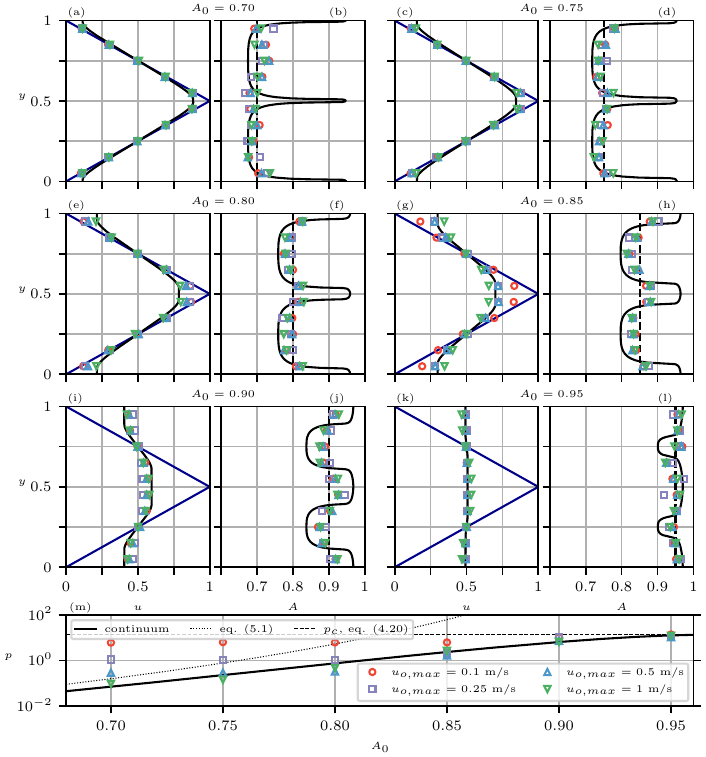}
	\caption{Solutions to the non-dimensional continuum model \eqref{eq:one_dim_system_normal} (black lines) compared with results from the DEM (markers), with the ocean velocity in blue. For each pair of panels in the first three rows, we fix the global sea ice concentration $A_0$, vary the maximum ocean velocity $u_{o,max}$ for the DEM, and plot the velocities in the panels to the left ((a), (c), (e), (g), (i), and (k)) and the local concentration to the right ((b), (d), (f), (h), (j), and (l)). In panel (m), we plot the pressure $p$ in terms of $A_0$. Due to the normalization in terms of $[u] = u_{o,max}$, the solutions to the continuum model are indifferent to a change in $u_{o,max}$.}
	\label{fig:comparison_fitted}
\end{figure}

\subsection{Variation in global concentration $A_0$ and ocean speed $u_{o,max}$}

We first evaluate the continuum model's accuracy in replicating the DEM results used for fitting the functions $\mu(I)$ and $\Phi(I)$ in section \ref{sec:inferring}. These results are computed for the ocean velocity profile \eqref{eq:uo_hat} and an ice thickness of $H = 2\,\si{m}$. We consider six global sea ice concentrations $A_0$ between $0.7$ and $0.95$ and four ocean velocities $u_{o,max}$ between $0.1$ and $1\,\si{m}/\si{s}$. For each of these cases, we run the DEM until a steady state is reached and then we extract 10 values of the sea ice velocity and concentration along the $y$-direction, as explained in section \ref{sec:inferring}, and one value for the pressure $p$, averaged over the whole square patch of ocean. 

Figure \ref{fig:comparison_fitted} displays the velocity and sea ice concentration profiles $u$ and $A$ and the pressure $p$ for both the continuum model and the DEM. The velocity profile $u$ and the pressure $p$ are normalized as explained in section \ref{subsec:nondim} using $[u] = u_{o,max}$. A consequence of this normalization is that the non-dimensional solutions $u$, $A$, and $p$ to the continuum model \eqref{eq:one_dim_system_normal} are indifferent to the value of $u_{o,max}$. For most of the DEM results, this is also the case. For each value of $A_0$, almost all of the normalized velocity profiles (panels (a), (c), (e), (g), (i), and (k)) and pressure points (panel (m)) appear to collapse onto a single curve or point, which is well approximated by the continuum model. 

\begin{figure}
	\centering
	\includegraphics[scale=1]{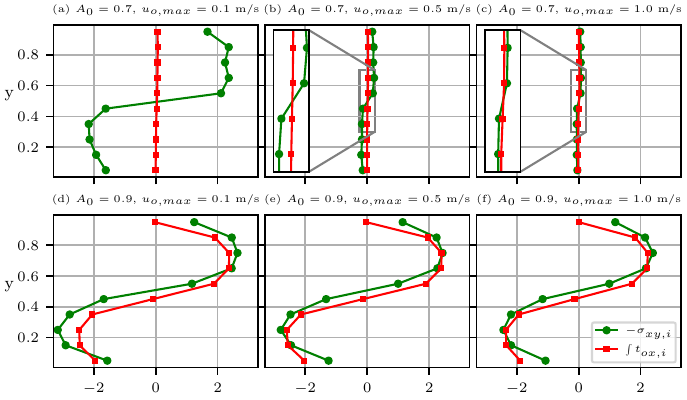}
	\caption{Shear stress $\sigma_{xy,i}$ and integrated ocean drag $\int t_{ox,i} := \Delta y \sum_{j = 1}^i t_{ox,j}$ extracted from the DEM for $A_0 = 0.7$ and $0.9$ and $u_{o,max} = 0.1$, 0.5 and $1\,\si{m/s}$.}
	\label{fig:force_balance_DEM}
\end{figure}

Departures from the other normalized DEM results are most visible for slow ocean currents and low concentrations. This becomes particularly clear when looking at the pressure in panel (m) when $u_{o,max} = 0.1\, \si{m}/\si{s}$ and $A_0 \leq 0.8$, where the pressure values from the DEM (circles) depart substantially from the prediction of the continuum model (black solid line). These mismatches signal the continuum model's limitations to capture the DEM results for the range of regimes considered. When fitting the dilatancy function $\Phi(I)$, the largest misfit is also found for points of slow ocean currents and low concentration ($u_{o,max} = 0.1\, \si{m}/\si{s}$ and $A_0 \leq 0.75$), see panel (b) in figure \ref{fig:fit}. If the DEM results are to approximate a continuum model as in \eqref{eq:one_dim_system_normal}, we expect the equality
\begin{align}\label{eq:force_balance}
	\sigma_{xy}(y) = - \int_0^y t_{ox}\,\dd y,
\end{align}
to hold approximately for the DEM results, with $t_{ox}$ denoting the horizontal component of the ocean drag. If we denote by $\sigma_{xy,i}$ and $t_{ox,i}$ the values extracted from the DEM at the grid cell of width $\Delta y$ located at $y_i$ (see appendix \ref{app:averaging} for an overview of how these quantities are obtained), a discrete balance analogous to \eqref{eq:force_balance} is 
\begin{align}\label{eq:force_balance_DEM}
	\sigma_{xy,i} = - \Delta y \sum_{j = 1}^i t_{ox,j}.
\end{align}
We remark that the term $t_{ox,j}$ results from averaging the ocean drag on each ice floe, as opposed to introducing the averaged sea ice velocity into \eqref{eq:def_to}. We also note that \eqref{eq:force_balance_DEM} ignores any choice of rheology and should hold independently of our choice of functions $\mu(I)$ and $\Phi(I)$. A surprising result we find when investigating the DEM data is that the terms to the left and right hand side of the equality in \eqref{eq:force_balance_DEM} differ in orders of magnitude whenever the sea ice concentration is low and the ocean currents are slow, see panel (a) of figure \ref{fig:force_balance_DEM}. Conversely, for faster ocean currents and denser concentrations, these terms become approximately equal (see panels (c)-(f) in figure \ref{fig:force_balance_DEM}). Inertial effects are found to be negligible in all of the cases we consider; moreover, since DEM quantities are averaged along the whole length of the domain in the $x$-direction, the term corresponding to $\partial \sigma_{xx}/\partial x$, which should be considered in a two-dimensional setting, becomes zero. Therefore, this finding raises the questions of whether a fundamentally different continuum model should be used for low sea ice concentrations and slow ocean currents or if a continuum model is valid at all.

The DEM's local concentration $A$ is accurately captured by the continuum model for low sea ice concentrations (panels (b), (d), (f), and (h) in figure \ref{fig:comparison_fitted}). For $A_0 = 0.90$ and 0.95 (panels (j) and (l)), the continuum model overestimates the degree of dilatancy, although the general trends are visibly similar, with regions of higher concentration around $y=0$, $1/2$, and $1$, where the strain rates are lowest. Since an increase in $\Delta$ diminishes the local variations in $A$ (see figure \ref{fig:solutions_compete_system}), this suggests that $\Delta$ could be adjusted to improve the fit with the data. In this case, it would enter the model as a physical parameter whose interpretation should be explored further. 

In panel (m) of figure \ref{fig:comparison_fitted}, we also test two limit approximations of the pressure which we find in section \ref{sec:analysis} when $u_o$ is given by \ref{eq:uo_hat_normal}. When $p \ll 1$, which occurs for the smaller values of $A_0$, we expect $\Cc(p) \approx 2^\alpha$ because $u \approx u_o$. This implies that
\begin{align}\label{eq:p_lim_0}
	p \approx 4 \frac{A_0}{n}\left(\frac{\phi_0}{1 - A_0}\right)^{2/\alpha} \quad \text{as $A_0 \to 0$}.
\end{align} 
On the other hand, high values of $A_0$ result in $I \ll 1$, see figure \ref{fig:fit}, and therefore $\mu(I) \approx \mu_0$, leading to the purely plastic regime studied in section \ref{subsubsec:plastic}. Panel (k) shows that the velocity profiles are mostly flat in this region, suggesting that the critical pressure $p_c$ calculated in \eqref{eq:pcritical} is a good approximation of the $p$. By plotting \eqref{eq:p_lim_0} and $p_c$ in panel (m), we find that these are indeed good approximations in their respective limits.

\begin{figure}
	\centering
	\includegraphics[scale=1]{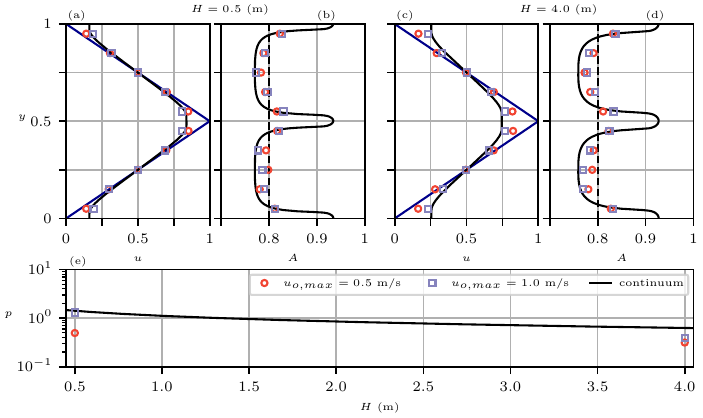}
	\caption{Solutions to the non-dimensional continuum model \eqref{eq:one_dim_system_normal} (black lines) compared with results from the DEM (markers). For each pair of panels in the first row, we fix the sea ice thickness $H$ and vary the maximum ocean velocity $u_{o,max}$ (the ocean velocity is plotted in blue in panels (a) and (c)). In panel (e), we plot the pressure $p$ in terms of $H$. Due to the normalization in terms of $[u] = u_{o,max}$, the solutions to the continuum model are indifferent to a change in $u_{o,max}$.}
	\label{fig:comparison_H}
\end{figure}

\subsection{Variation in ice thickness and number of floes}

Next, we test the effectiveness of the continuum model in capturing the DEM results for different ice thicknesses and floe sizes. In figure \ref{fig:comparison_H}, we plot results for the DEM and the continuum model for steady states with $H= 0.5$ and $H = 4\,\si{m}$, different ocean velocities $u_{o,max}$, and a global sea ice concentration of $A_0 = 0.8$. The ice thickness $H$ enters the continuum model via the parameter $\epsilon = H/L$ in the momentum equation \eqref{eq:mom_normal}. An increase in $H$ is accompanied by an increase in $\epsilon$, which effectively acts as a decrease in the normalized external forcing. We therefore expect an increase in $H$ to result in a decrease in the shear strain and pressure. As observed in panels (a) and (c) of figure \ref{fig:comparison_H}, this is the case for the velocity profiles of both the continuum model and the DEM; an increase in $H$ is accompanied by a flattening of the normalized velocity profiles. The continuum model is once again accurate in capturing the averaged velocities of the DEM and the general trends in the variation of the sea ice concentration. The pressure resulting from the continuum model and the DEM also decreases as the ice thickness increases, see panel (f); for the pressure, we find a decent fit between the DEM and the continuum model.

\begin{figure}
	\centering
	\includegraphics[scale=1]{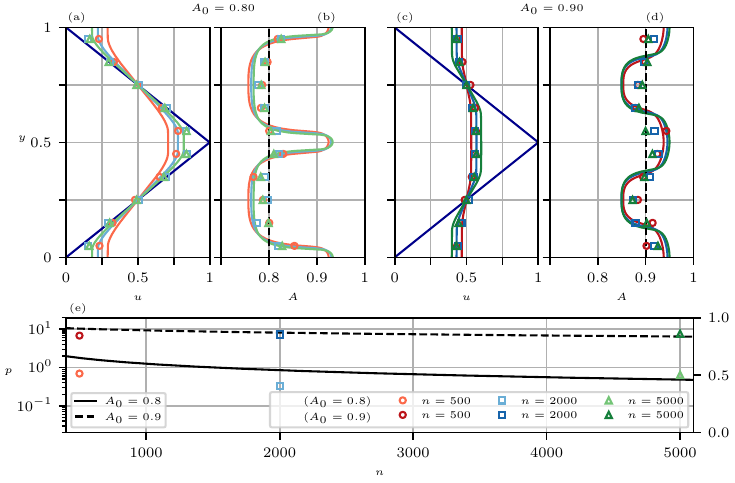}
	\caption{Solutions to the non-dimensional continuum model \eqref{eq:one_dim_system_normal} (lines) compared with results from the DEM (markers). The solutions to the continuum model are plotted in red for $n = 500$, blue for $n = 2000$ and green for $n = 5000$; the tone is set to light if $A_0 = 0.8$ and dark if $A_0 = 0.9$ (see legend in (e)). For each pair of panels in the first row, we fix the sea ice thickness $H$ and vary the maximum ocean velocity $u_{o,max}$ (the ocean velocity is plotted in blue in panels (a) and (c)). In panel (e), we plot the pressure $p$ in terms of $n$.}
	\label{fig:comparison_n}
\end{figure}

Finally, figure \ref{fig:comparison_n} contains a comparison between the DEM and the continuum model for different numbers of floes $n$. In particular, we present results with $n = 500$, 2000 and 5000 for $A_0 = 0.8$ and $0.9$ and $u_{o,max} = 0.5\,\si{m}/\si{s}$. An increase in $n$ implies a decrease in the effective viscosity of the model, and we therefore expect a decrease in the ice strength or pressure $p$; in the limit where $n\to\infty$, the rheology becomes purely plastic, this can be seen by taking this limit in \eqref{eq:mom_normal}. In figure \ref{fig:comparison_n}, one can see that the pressure in continuum model indeed decreases with $n$, but this is not the case with the DEM. In fact, the results of the DEM appear to change little with $n$, especially for $A_0 = 0.9$. Despite this, the results from the DEM do not depart from those of the continuum model excessively. 

\section{Comparisons with existing models for sea ice}\label{sec:similarities_model}

The continuum model studied in this paper shares features with existing models for sea ice. Here, we examine those similarities and also establish differences with our continuum model. Due to its ubiquity in sea ice modeling, we first consider Hibler's model in section \ref{subsec:hibler}, before considering other models in section \ref{subsec:other-models}.

\subsection{Hibler's model}\label{subsec:hibler}

The most widely used model for sea ice is Hibler's model, which was first presented in \citet{hibler1979} and treats sea ice like a viscoplastic material. Under the one dimensional conditions of the steady square ocean patch and the non-dimensionalization in section \ref{sec:analysis}, Hibler's model reduces to the following form:
\begin{subequations}\label{eq:hibler-model}
\begin{align}
	- \epsilon \frac{\dd }{\dd y}\left( \frac{p}{2e} \frac{\dd u/\dd y}{\sqrt{|\dd u/\dd y|^2 + \delta^2}} \right)= \beta_o |u_o - u|(u_o - u),\label{eq:hibler-mom}\\
	p = \frac{P^\ast}{\rho_iu_{o,max}^2}\exp{(-20(1-A))}. \label{eq:hibler-p}
\end{align}	
\end{subequations}
In \eqref{eq:hibler-model}, the parameter $P^\ast$ is an empirical constant, $\delta$ is a regularization parameter, and $e = 2$ represents the eccentricity of the elliptical yield curve from which Hibler's model is derived. That is, Hibler's model assumes a plastic rheology based on an elliptical yield curve which is then regularized. Inside the plane of principal stresses, the yield curve is set in such a way that the ice only resists compression, not pure extension. This geometrical configuration is motivated by the observation that, in pack ice, deformation occurs through ridging (compression) and the opening of leads (extension); of these two mechanisms, only ridging requires a non-negligible amount of plastic work \citep{rothrock1975}.

In the steady one dimensional setting, equation \eqref{eq:hibler-mom} can be recovered from the regularized momentum equation \eqref{eq:mom_reg} by setting $\mu_0 = 1/(2e)$ and $\mu_1 = 0$ (pure plasticity). In fact, $1/(2e) = 0.25$, which comes very close to the value $\mu_0 = 0.26$ which we infer from the DEM. This means that close to the quasi-static regime, when $I \ll 1$ and $\mu(I) \approx \mu_0$, we essentially work with the momentum equation from Hibler's model. This is a very reasonable coincidence, because Hibler's model was designed for the central ice pack where the sea ice concentration is very high. 

\begin{figure}
	\centering
	\includegraphics[scale=1]{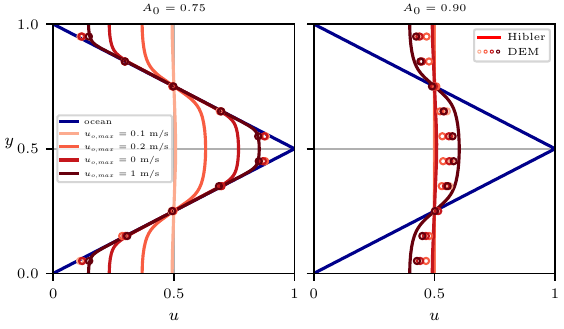}
	\caption{Solutions to Hibler's model for different global concentrations $A_0$ and ocean velocities $u_{o,max}$. We also plot the results from the DEM as a reference. Here, we set $H = 2\,\si{m}$ and $n = 2000$.}
	\label{fig:hibler}
\end{figure}

The main departure between our model and Hibler's is the expression for the pressure \eqref{eq:hibler-p}: as $u_{o,max}$ increases, $p$ decreases in Hibler's model, unlike our continuum model where $p$ is independent of $u_{o,max}$. This makes it impossible for Hibler's model to capture the invariance of the non-dimensional sea ice velocity and pressure with $u_{o,max}$ which we find in most of the DEM results in section \ref{sec:comparison_DEM}. This is made clear in figure \ref{fig:hibler}, where we show solutions to Hibler's model and compare them with the DEM for different values of $A_0$ and $u_{o,max}$; an increase in $u_{o,max}$ is accompanied by excessively large changes in the velocity profile. We set $P^\ast = 5\times 10^4\,\si{N}\si{m}^{-1}$ and $\delta = 0.1$ n order to get a good fit with the DEM for some values of $u_{o,max}$. We remark that Hibler's model is designed for ice that fractures and ridges; in this setting, an increase in the ocean drag weakens the ice through this mechanical deformation. Below, in section \ref{sec:conclusions}, we state that future extensions of our continuum model should include the effects of fracturing and ridging, available in SubZero. These future investigations should study the validity of \eqref{eq:hibler-p} when fracturing and ridging are accounted for.

The choice of an elliptical yield curve in Hibler's model is motivated by the simplicity of the resulting rheological formulation, yet other possibilities consistent with the requirement of null resistance to pure extension are available, such as the parabolic lens and the teardrop yield curves \citep{zhang2005, ringeisen2023} and the Mohr-Coulomb criterion \citep{ip1991}. In fact, as mentioned at the end of section \ref{sec:inferring}, the purely plastic rheology in the $\mu(I)$ formulation considered here (given by \eqref{eq:rheology_2D} with $\mu_1 = 0$), can be written as the Mohr-Coulomb rheology presented in \citet{ip1991} and \citet{gutfraind1997}. This becomes clear if we note that $\|\bS\| = \mathrm{D}_1 - \mathrm{D}_2$, where $\mathrm{D}_1$ and $\mathrm{D}_2$ denote the principal components (eigenvalues) of $\bD$, and we compare the plastic component in \eqref{eq:rheology_2D} with expression (6) in \citet{gutfraind1997}. A significant difference between the two expressions is that in (6) from \citet{gutfraind1997} the viscosity is capped to a maximum value in order to avoid the inevitable blow-up that occurs with a purely plastic rheology; this is another form of regularization. This difference of plastic yield curves between Hibler's model and the purely plastic version of our model is another point of departure between both models in a two dimensional setting.

\subsection{Other models}\label{subsec:other-models}

In section \ref{sec:inferring}, we explain that the viscous component in \eqref{eq:rheology_2D} can be interpreted as the rheological effects of collisions, which become increasingly important as $I$ increases. As mentioned in that section, a collisional rheology is derived in \cite{shen1987} which is also of a linearly viscous nature. In fact, this collisional rheology establishes a viscosity which is linearly proportional to $\overline{d} \sqrt{\rho_i H p}$, as in \eqref{eq:rheology_2D} (see \citet{herman2022} for a clear description of the collisional rheology). For this, the model proposed in \citet{feltham2005} comes close to \eqref{eq:rheology_2D}, since it takes the ice rheology to be the sum of Hibler's rheology and the collisional rheology from \citet{shen1987}.

To the knowledge of the authors, the only other study using the $\mu(I)$ rheology to model sea ice is due to \cite{herman2022}. There, the floes are driven by a moving wall (as opposed to an ocean or atmospheric current, as in our case), and the DEM is based on disk shaped floes, with a much more severe polydispersity. Two main differences can be found between the function $\mu(I)$ that we infer (figure \ref{fig:fit}) and the one found in \cite{herman2022}. First, although both cases consider a very similar range of $I$ values, the magnitude of the friction $\mu$ differs considerably, although it is of the same order of magnitude. Second, the $\mu(I)$ curve in \cite{herman2022} plateaus for $I > 10^{-1}$, something that we do not observe. It remains unclear what may cause these differences. Regarding the second point, we justify our use of a linear function for $\mu$, as in \eqref{eq:fitting_functions}, by remarking that it simplifies the resulting constitutive equation, enabling a more thorough analysis of the model, and resembles the $\mu$ function proposed in \cite{herman2022} over a large range of $I$ values.

\section{Conclusions and future work}\label{sec:conclusions}

In this paper, we have presented a novel mathematical model for sea ice in the MIZ based on the $\mu(I)$ rheology. We have inferred the form of this rheology in section \ref{sec:inferring} using data produced with DEM computations. With the analysis in section \ref{sec:analysis}, we prove that the steady one dimensional formulation of this model, given by \eqref{eq:one_dim_system_normal}, is well-posed in the sense that it has a unique solution. The numerical results presented in section \ref{sec:comparison_DEM} demonstrate that this model is capable of replicating most of the results of the DEM in the context of steady one dimensional problems. The most visible departure from the continuum model occurs for low ocean velocities and sea ice concentrations, with $u_{o,max} = 0.1\,\si{m}/\si{s}$ and $A_0 \leq 0.85$; in this case, the DEM results indicate that the shear stress no longer balances the integrated ocean drag, signaling a breakdown of the underlying grid-averaged momentum balance equation. That is, since inertial effects are found to be negligible in the steady states we consider and the extensional stresses cancel out when integrating across the $x$-direction of the domain to perform the averaging of DEM quantities, our basic continuum model, prior to any choice of rheology, establishes a balance between shear stresses and ocean drag. In figure \ref{fig:force_balance_DEM}, this balance is found to approximately hold for all cases but those of slow ocean currents and low sea ice concentrations.

The lack of validity of the continuum model for slow ocean currents and low concentrations found in section \ref{sec:comparison_DEM} should be explored further, since this is a regime we expect to find in parts of the MIZ \citep{stewart2019}. As explained in section \ref{sec:comparison_DEM}, this lack of accuracy is accompanied by the breakdown of the balance between the averaged shear stress and the integrated ocean drag extracted from the DEM. This balance precedes the choice of rheology and therefore indicates that either a fundamentally different continuum model should be used or some assumption necessary for a continuum model to even hold is no longer valid.

Mechanical processes like fracturing, ridging, and welding are fundamental processes in sea ice dynamics \citep{feltham2008}. Therefore, future improvements of the model presented here should consider the inclusion of these effects. Since ridging and fracturing are fundamental processes in the central ice pack \citep{rothrock1975}, our continuum model could also be valid in this area when these effects are included, yielding a unified sea ice model. Since SubZero includes these mechanical interactions between ice floes \citep{manucharyan2022}, it is a promising tool for exploring their macroscopic effects on the rheology. A first step could be to explore the consequences of including floe fracturing. We expect this would set an upper bound on the pressure and shear stress that the ice cover can sustain. This bound will probably be closely related to the fracture criterion used at the ice floe level. Moreover, floe fracturing will create smaller floes that lead to higher degrees of polydispersity. 

The comparisons between the continuum model and the DEM in section \ref{sec:comparison_DEM} are confined to the context of the steady periodic ocean problem, which is essentially one dimensional. Any future studies must examine the accuracy of the $\mu(I)$ rheology in modeling sea ice dynamics under unsteady conditions and in a two-dimensional configuration. \citet{barker2015} discovered the $\mu(I)$ formulation presented in section \ref{subsec:2D} to be mathematically ill-posed in time-dependent problems, and \cite{schaeffer2019} has proposed modifications that avoid these instabilities while leaving the steady equations unchanged. A future investigation in the context of sea ice should take these studies into account. 

The rheology we propose in \eqref{eq:rheology_2D} is local in the sense that the viscosity and the pressure at a certain point of the domain only depend on other quantities and their derivatives at that same point. Yet, granular materials create complex contact networks that enable the interaction of grains set far apart. This leads to non-local effects, and extensions of the $\mu(I)$ rheology that include these effects have been proposed in the context of granular flows \citep{kamrin2012, bouzid2013}. Such effects will probably also arise in sea ice modeling and should be considered in future extensions of the model we propose here.

\appendix

\section{Some notes on the DEM SubZero}

\subsection{Computing the stress tensor for an ice floe}\label{app:collision}

At a given instant in time, the floe $i$ is in contact with floes whose indices are given by the set $C_{i}$. For each floe $j\in C_{i}$, there are $n^c_{i,j}$ contact points (there can be several contact points between two floes if one of them is concave). The stress tensor $\bsigma_i$ of floe $i$ is given by
\begin{align}\label{eq:sigmai}
	\bsigma_i = \frac{1}{a_i}\sum_{j\in C_i}\sum_{k = 1}^{n^c_{i,j}}  \bff^k_{i,j} \otimes \br^k_{i,j},
\end{align}
where $a_i$ is the area of floe $i$, $\bff^k_{i,j}$ is the force at the $k$th contact point exerted by floe $j$ on floe $i$, and $\br^k_{i,j}$ is the vector connecting the center of mass of floe $i$ with the $k$-th contact point. Expression \eqref{eq:sigmai} corresponds with the Love-Weber formula and, in general, additional terms corresponding to dynamics effects must also be accounted for \citep{nicot2013}. However, we find these dynamic terms to be negligible in all of our computations. We also remark that \eqref{eq:sigmai} differs from its counterpart in \citep[equation (9)]{manucharyan2022} in two points: (1) we divide by the floe area $a_i$ rather than its volume $a_iH_i$ to obtain the right units and account for the fact that we are working with depth-integrated stresses. (2) We avoid forcing $\bsigma_i$ to be symmetric and simply use the Love-Weber formula. In general, we find that $\sigma_{i,xy} \approx \sigma_{i,yx}$ in all of our DEM computations.

\begin{figure}
	\centering
	\includegraphics[scale=1]{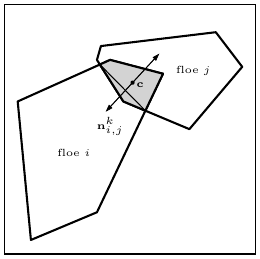}
	\caption{Collision between two floes. The contact forces are calculated in terms of geometric properties of the overlap area, shown in gray. Here, $\bc$ represents the center of mass of the overlap area and the normal direction $\bn^k_{i,j}$ is taken normal to the line connecting the two intersection points between the floes.}
	\label{fig:floe-collision}
\end{figure}

For the convenience of the reader, we now summarize the calculation of contact forces between two colliding floes in SubZero. A complete account is given in \citep{manucharyan2022}. The contact force $\bff^k_{i,j}$ is the sum of its normal and tangential components,
\begin{align}
	\bff^k_{i,j} = \bff^{N,k}_{i,j} + \bff^{T,k}_{i,j}.
\end{align}
In figure \ref{fig:floe-collision} we represent the parameters and vectors involved in the collision of two floes. Two colliding floes intersect; this intersection results in an overlap polygon, represented in grey in figure \ref{fig:floe-collision}, of area $\Ac$ and center of mass $\bc$. The point $\bc$ is then considered the contact point between floes $i$ and $j$. The normal direction $\bn^k_{i,j}$ from floe $j$ to $i$ is defined perpendicular to the chord uniting the two intersection points between both floes, as in figure \ref{fig:floe-collision}. The normal force is then defined as 
\begin{align}
	\bff^{N,k}_{i,j} = \kappa \Ac \bn^k_{i,j},
\end{align}
where 
\begin{align}\label{eq:kappa}
	\kappa = E \frac{H_i H_j}{H_i d_i + H_j d_j}.
\end{align}
In \eqref{eq:kappa}, $E$ is Young's modulus, $H_i$ the floe thickness, and $d_i = \sqrt{a_i}$ a measure of the floe size. Normal forces are thus elastic and do not dissipate energy.

The tangential force is given by 
\begin{align}
	\bff^{T,k}_{i,j} = c_{i,j}^k G\, \Delta t\, v_{i,j}^k |\bff^{N,k}_{i,j}| \bt_{i,j}^k,
\end{align}
where $c_{i,j}^k$ is the length of the chord uniting the two intersection points between floes $i$ and $j$ and $G$ is the shear modulus $G = E/(2(1 + \nu))$, defined in terms of Poisson's ratio $\nu$. The parameter $\Delta t$ is the simulation's time step, $v_{i,j}^k$ the tangential velocity difference between both floes, and $\bt_{i,j}^k$ the tangential direction. The tangential force is limited to the following upper bound:
\begin{align}
	|\bff^{T,k}_{i,j}| \leq \mu^\ast |\bff^{N,k}_{i,j}|,
\end{align}
where $\mu^\ast$ is the inter-floe friction coefficient.

\subsection{Spatial averaging of data}\label{app:averaging}
\begin{figure}
	\centering
	\includegraphics[scale=1]{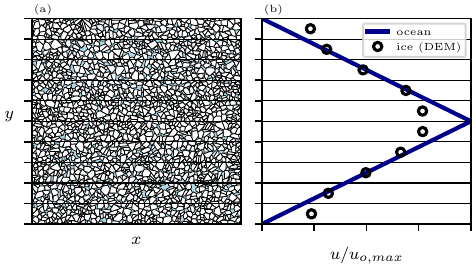}
	\caption{The values of the horizontal velocity $u$ and the stress tensor $\bsigma$ are extracted from the DEM data for each time step by averaging spatially over the 10 regions, elongated in the $x$-direction, shown here. In this figure, $n = 2000$, $u_{o,max} = 2\,\si{m}/\si{s}$, and $A_0 = 0.8$.}
	\label{fig:inference}
\end{figure}

To average the DEM data in space, we must grid the square domain. Taking into account the one dimensional nature of the problem, we divide the domain into $N = 10$ cells that stretch in the $x$-direction, such that the edges separating these cells are defined along $N+1$ equispaced points $(y_0, y_1, ..., y_{N})$ in the  $y$-direction. Hence, the resulting grid consists in the cells $[0,L]\times[y_{i-1},y_{i}]$ for $i=1,...,N$, as depicted in figure \ref{fig:inference}. At each time step, we compute the horizontal velocity $u$ and the components of the stress tensor $\bsigma$ by spatially averaging the velocities and stresses of the ice floes contained in each region (the manner in which the stress tensor $\sigma$ is computed for each ice floe is explained in above in appendix \ref{app:collision}. In particular, for each cell of the grid, we perform a mass-weighted averaging such that, for ice floes that are only partially contained in the cell, only the mass of the floe inside the cell is considered. More information about the averaging can be found in \citet{manucharyan2022}. In order to be consistent with \eqref{eq:one_dim_system_mom_y}, for each run we extract a single pressure $p = \frac{1}{2}(\sigma_{xx} + \sigma_{yy})$ by spatially averaging this quantity over the whole domain.

By performing the spatial and temporal averaging, for each DEM computation we obtain a pressure $p\in \RR$ and the vectors $(u_i)$, $(\sigma_{xy,i})$, and $(A_i)$ of data points in $\RR^N$. To compute the inertial numbers $I$, we first find the strain rate vector $(\dd u_i)$ via central finite differences, such that $\dd u_i = (u_{i+1} - u_{i-1})/(2(y_{i+1} - y_{i-1}))$. Then, $I_i = \overline{d}\sqrt{H\rho_i/p}|\dd u_i|$, with $\overline{d} = \sqrt{Ao L^2/n}$.

\section{Existence and uniqueness of solutions to the momentum equation}\label{app:existence}

Since the momentum equation \eqref{eq:mom_no_sing} of the continuum model is equivalent to the minimization of the functional $\Jc$ over the space $V$, defined in \eqref{eq:def_V}, it suffices to show that $\Jc$ admits a unique minimizer to prove the existence and uniqueness of solutions to the momentum equation. To do so, we first write the functional $\Jc:V\to \RR$ as follows for simplicity:
\begin{align}
	\Jc(v) = \gamma_1 \left|v\right|_1 + \gamma_2 \left|v\right|^2_2 + \gamma_3 \| u_o - v \|_3^3,
\end{align}
where $\gamma_i > 0$ for $i = 1,2,3$ are constants and, for $q \geq 1$, $\|\cdot\|_q$ and $\left|\cdot\right|_q$ are the Sobolev norm and semi-norms, respectively, for the $L^q((0,1))$ and $W^{1,q}((0,1))$ spaces; more precisely, these are defined by
\begin{align}
	\|v\|_q = \left(\int_0^1 \left|v\right|^q\,\dd x\right)^{1/q} \quad \text{and} \quad \left|v\right|_q = \left(\int_0^1 \left|\frac{\dd v}{\dd x}\right|^q\,\dd x\right)^{1/q}.
\end{align}
We first remark that $\Jc$ is well defined for all $v\in V$ by the Sobolev embedding theorem, which implies that $ \| v \|_3 < \infty$. Moreover, for the sake of rigor, we must also assume that $u_o \in L^3((0,1))$.

According to \citep[theorem 2, chapter 8]{evans2010}, at least one function $V$ exists that minimizes $\Jc$ if the functional is convex and coercive. It is straightforward to check that $\Jc$ is convex. The functional $\Jc$ is said to be coercive in $V$ if 
\begin{align}\label{eq:coercivity}
	\|v\|_V \to \infty \implies \Jc(v) \to \infty,
\end{align}
where $\|v\|^2_V = \|v\|_2^2 + |v|_2^2$. We first note that, by the triangle inequality and Young's inequality, it follows that
\begin{align}
	\|v\|^3_3 \leq 4\left( \|u_o - v\|^3_3 + \|u_o\|^3_3\right).
\end{align}
Therefore, for any $v\in V$,
\begin{align}
	\Jc(v) \geq \gamma_1 \left|v\right|_1 + \gamma_2 \left|v\right|^2_2 + \frac{\gamma_3}{4}\| v \|_3^3 - \gamma_3 \|u_o \|_3^3
\end{align}
Using H\"older's inequality we can show that $\|v\|_3 \geq \|v\|_2$, such that
\begin{align}
	\Jc(v) \geq \gamma_2 \left|v\right|^2_2 + \frac{\gamma_3}{4}\| v \|_2^3 - \gamma_3 \|u_o \|_3^3,
\end{align}
from where coercivity follows. We note that this argument fails whenever $\gamma_2 = 0$, which corresponds with the purely plastic regime, because $\left|v\right|_1 \to \infty$ does not necessarily follow from $\|v\|_V \to \infty$.

To prove that there is only one function that minimizes $\Jc$, we assume by contradiction that both $u_1$ and $u_2$ minimize $\Jc$ and $u_1\neq u_2$. In this case, we must have that $\Jc(u_1) = \Jc(u_2)$. We define $w = 1/2(u_1 + u_2)$ and note that
\begin{align}
	\|u_o - w \|_3^3 < \frac{1}{2}\|u_o - u_1 \|_3^3 + \frac{1}{2}\|u_o - u_2 \|_3^3,
\end{align} 
due to the strict convexity of the function $\|\cdot\|^3$ in $\RR$ and the injectivity of $\|u_o - \cdot \|_3^3$ in $V$. As a result, by appealing to the convexity of the seminorm $\|\cdot\|_q$ for all $q \geq 1$,
\begin{align}
	\Jc(w) < \frac{1}{2} \Jc(u_1) + \frac{1}{2} \Jc(u_2) = \Jc(u_1),
\end{align}
a contradiction because $\Jc(u_1) \leq \Jc(v)$ for all $v\in V$. 

\backsection[Acknowledgements]{We thank Andrew Thompson for many insightful comments and suggestions shared with us during the preparation of the manuscript.} 

\backsection[Funding]{This work has been supported by the Multidisciplinary University Research Initiatives (MURI) Program, Office of Naval Research (ONR) grant number N00014-19-1-242.}

\backsection[Declaration of Interests]{The authors report no conflict of interest.}

\backsection[Author contributions]{All authors contributed to conceiving and designing the study. GD wrote the paper, designed the figures, performed the simulations and mathematical analysis. MG, SG and GS revised the paper. MG, RH and SG contributed to the simulations, and SG contributed simulation tools. MG, GS and RH provided input to the analysis.}

\bibliographystyle{jfm}
\bibliography{bibliography}

\end{document}